\documentclass[conference]{IEEEtran}
\ifCLASSINFOpdf
\else
\fi
\usepackage{fancyhdr}

\usepackage[sort,nocompress]{cite}
\usepackage[hyphens]{url}
\usepackage{graphicx}
\usepackage{textcomp}
\usepackage{xcolor}
\usepackage[bookmarks=true,breaklinks=true,letterpaper=true,colorlinks,linkcolor=black,citecolor=blue,urlcolor=black]{hyperref}

\usepackage[normalem]{ulem}

\usepackage{soul}
\usepackage{algorithm}
\usepackage{xspace}
\usepackage[font=footnotesize,labelfont=bf]{caption}
\usepackage{mathtools}
\usepackage{amsfonts}
\usepackage{amssymb}
\usepackage{pifont}
\usepackage{color}
\usepackage{algpseudocode}
\usepackage{multirow}
\usepackage{textgreek}
\usepackage{subfig}

\newcommand{\fig}[1]{Figure~\ref{#1}}
\newcommand{\sect}[1]{Section~\ref{#1}}
\newcommand{\tab}[1]{Table~\ref{#1}}
\newcommand{\algo}[1]{Algorithm~\ref{#1}}
\newcommand{\eqn}[1]{Equation~\ref{#1}}

\algnewcommand{\LineComment}[1]{\State \(\triangleright\) #1}
\newcommand{\elsa}[0]{ELSA\xspace}
\newcommand{\paris}[0]{PARIS\xspace}
\newcommand{\knee}[0]{\emph{MaxBatch$_{knee}$}\xspace}

\newcommand{\Small}[1]{GPU(#1)\xspace}
\newcommand{\Medium}[1]{GPU(#1)\xspace}
\newcommand{\Largest}[1]{GPU(#1)\xspace}

\newcommand\blfootnote[1]{%
\begingroup
\renewcommand\thefootnote{}\footnote{#1}%
\addtocounter{footnote}{-1}%
\endgroup
}

\def\BibTeX{{\rm B\kern-.05em{\sc i\kern-.025em b}\kern-.08em
    T\kern-.1667em\lower.7ex\hbox{E}\kern-.125emX}}

\renewcommand{\baselinestretch}{.999}
\title{PARIS and ELSA: An Elastic Scheduling Algorithm\\ for Reconfigurable Multi-GPU Inference Servers}

\author{
	\IEEEauthorblockN{
		Yunseong Kim,
		Yujeong Choi,
		Minsoo Rhu}

	\IEEEauthorblockA{
		School of Electrical Engineering \\
		KAIST}

	\IEEEauthorblockA{
		\texttt{\{yskimno1\}}@gmail.com, \texttt{\{yjchoi0606, mrhu\}}@kaist.ac.kr}
}

\author{
	\IEEEauthorblockN{
		Yunseong Kim,
		Yujeong Choi,
		Minsoo Rhu}

	\IEEEauthorblockA{
		School of Electrical Engineering \\
		KAIST}

	\IEEEauthorblockA{
		\texttt{\{yskimno1, yjchoi0606, mrhu\}}@kaist.ac.kr}
}

\begin{document}

\maketitle
\pagestyle{plain}

\begin{abstract}

In cloud machine learning (ML) inference systems, providing low latency
to end-users is of utmost importance. However, maximizing server
utilization and system throughput is also crucial for ML service
	providers as it helps lower the total-cost-of-ownership.  GPUs have oftentimes
been criticized for ML inference usages as its massive compute and memory
throughput is hard to be fully utilized under low-batch inference scenarios. To
address such limitation, NVIDIA's recently announced Ampere GPU architecture provides
features to ``reconfigure'' one large, monolithic GPU into multiple smaller ``GPU
partitions''. Such feature provides cloud ML service providers the ability to
utilize the reconfigurable GPU not only for large-batch training but also
for small-batch inference with the potential to achieve high resource utilization.
In this paper, we study this emerging GPU architecture with reconfigurability
to develop a high-performance multi-GPU ML inference server. Our first proposition
is a sophisticated partitioning algorithm for reconfigurable GPUs that systematically
determines a heterogeneous set of multi-granular GPU partitions,
					 best suited for the inference server's deployment. Furthermore, we co-design
					 an elastic scheduling algorithm tailored for our heterogeneously partitioned
					 GPU server which effectively balances low latency and high GPU utilization.

\end{abstract}

\IEEEpeerreviewmaketitle
\blfootnote{
This is an extended version of our work, which is accepted for publication at the $59^{th}$
	Design Automation Conference (DAC), $2022$.
}

\section{Introduction}
\label{sect:intro}

Several hyperscalers are now offering ``MLaaS (Machine Learning as a Service)''
from cloud datacenters using off-the-shelf CPUs, GPUs, or even custom designed
accelerators for ML~\cite{cloud_tpu,amazon_elastic_inference,habana_goya}.
For end-users utilizing MLaaS for inference, providing real-time response
with strict SLA (service-level agreement) guarantee is of utmost importance. From
a MLaaS provider's perspective however, achieving high server resource utility and
system throughput is crucial as it helps optimize the total-cost-of-ownership (TCO)
	of maintaining the consolidated/virtualized datacenter infrastructure.

	 Unlike the throughput-bound ML training algorithm, inference is a latency-sensitive workload
	 which favors inference purpose built ML
	 accelerators~\cite{habana_goya,tpu_paper,kingscanyon} or even
	 latency-optimized CPUs~\cite{hazelwood2018applied,gupta2020architectural}.
	 GPUs on the other hand have generally been considered ill-suited for
	 latency-critical inference servers as its massive computational throughput
	 and memory bandwidth is hard to be fully utilized under low-batch inference
	 scenarios. Indeed, multiple prior literature motivated the
	 need for inference-optimized ASIC/FPGA solutions~\cite{brainwave_isca,moss2018customizable,nurvitadhi2017can}, criticizing
		GPUs for its low ``effective'' throughput and low utilization when deployed for inference.
	 To address such limitation, NVIDIA's recently announced Ampere
	 architecture provides a feature named \emph{Multi-Instance GPU} (MIG) that
	 enables the compute and memory resources of one large GPU to be
	 \emph{reconfigured} into multiple small or medium sized ``\emph{GPU partitions}''. 
	 As the partitioned GPUs are virtualized  and can be
	 handed over to multiple VMs using hardware support for SR-IOV~\cite{sriov,tulloch2013optimizing}, each GPU partition can function as a standalone GPU with
	 performance isolation guarantees. 
	 Such feature
	 can come in handy for MLaaS providers as the
	  reconfigurable GPU can be utilized not only for training
	 (i.e., configured as one big GPU) but also for low-batch inference  with the potential to
	 achieve high resource utility (i.e., partitioned into multiple small/medium
			 sized GPUs that suits application's characteristics).

	Given such landscape, a key objective of our study is to study this emerging
	GPU architecture with reconfigurability to develop a high-performance
	multi-GPU ML inference server. We first start by characterizing the pros/cons
	of the reconfigurable GPU when statically partitioned into a
	\emph{homogeneous} set of fixed size small (or medium) GPUs. 
	Our characterization
	reveals several limitations of a homogeneously partitioned
	multi-GPU inference server.
	As we
	explore in this work, determining the optimal GPU partition size  requires
	careful consideration of not just the target ML application's unique
	compute/memory needs, but  also the input query size (i.e., batch size).
	However, tackling such multi-dimensional optimization problem via a
	``one-size-fits-all'' approach (i.e., blindly partitioning the reconfigurable
			GPU into a statically fixed granularity) is not practical as the system
	architect must painstakingly explore the wide design space of GPU reconfigurability, batch size, and DNN models altogether, leading to suboptimal
	design decisions and 
incurring 
	either significant SLA violations or GPU underutilization. 

To this end, we propose a sophisticated yet practical {\bf P}artitioning {\bf
	A}lgorithm for {\bf R}econfigurable multi-GPU {\bf I}nference {\bf S}ervers
	(\paris) that systematically determines a \emph{heterogeneous} set of
	multi-granular GPU partitions in a user-transparent manner, best suited for the inference server's
	deployment scenario. Compared to a statically partitioned homogeneous GPU
	inference server, \paris presents rich opportunities to minimize GPU
	underutility while still providing enough computation power to satisfy SLA.
	We also present an {\bf EL}astic {\bf S}cheduling {\bf
		A}lgorithm (\elsa), co-designed with our \paris,  which is capable of
		exploiting the unique heterogeneous compute capabilities of our proposed
		multi-GPU server for scheduling decisions, effectively balancing low
		latency and high GPU utilization.

\section{Background}
\label{sect:background}

\subsection{Training vs. Inference in Machine Learning}
\label{sect:training_vs_inference}

A deep neural network (DNN) application must first be \emph{trained} to be
ready for deployment in \emph{inference} use-cases. Under the context of
training, the input training dataset is readily available before the learning
process is initiated, so establishing a large enough \emph{input batch size} is
trivial (e.g., the input batch size for training can be up to several hundreds
		to even thousands of inputs per batch~\cite{megatron-lm,mudigere2021high,gshard}). In contrast, batching
multiple inputs for inference is challenging as the inference server receives
DNN inference queries at varying rates, a function determined by what time of
the day the queries are being received, how much popular the deployed service
is, and more. In general, several prior work observed that the input query
arrival rate for web-based services follow a Poisson distribution with the
query size (i.e., batch size) following a log-normal
distribution~\cite{mattson2020mlperf,deeprecsys,li2016work,barford1998generating}. A high-performance ML inference server must
therefore carefully consider both query arrival rate and query size
distributions and be provisioned with sufficient amount of compute and memory resources to
satisfy SLA.

\subsection{GPUs for Training vs. Inference}
\label{sect:gpu_effect}

GPUs have traditionally been optimized in a throughput-centric fashion,
		 employing an area-efficient SIMD-based many-core  architecture design
		 backed with bandwidth-optimized memory solutions like GDDRx or
		 HBM~\cite{gddr6,hbm}.  This is in stark contrast to latency-optimized CPUs
		 where the primary design objective is to minimize latency using
		 sophisticated branch predictors, prefetchers, large on-chip caches, etc.
		 Consequently, throughput-hungry ML training algorithms are well suited for
		 GPUs as it can provide much higher throughput (per area) vs.  CPUs.
		 Inference however is a latency-critical workload, favoring purpose built
		 ML accelerators optimized for latency or even CPUs over GPUs. As discussed in
		 \sect{sect:training_vs_inference}, the batch size of an inference query is
		 typically orders of magnitude smaller than those for training. As a
		 result, the resource demands of inference are generally
		 not high enough to fully saturate the massive compute/memory 
		 throughput of GPUs.  Inference servers therefore can significantly suffer
		 from low GPU utilization, making it a less favorable choice for
		 TCO-optimized datacenters.  

		 To remedy such situation, vendors have introduced several lightweight, inference-purposed GPUs to the market which are equipped with a (relatively) smaller compute
		 capability (e.g., NVIDIA M4/T4~\cite{m4,t4}).  Employing these \emph{small} GPUs for
		 inference servers however has an important tradeoff as it reduces the compute ``density''
		 of the inference server, proportional to the performance difference between
				 large vs. small GPUs. Recently announced GPUs therefore are architected
		 with ``reconfigurability'' that enables them to be setup as one large, monolithic GPU or be \emph{partitioned} into multiple smaller GPUs, the granularity of which can be chosen by system architects as appropriate per application needs. Below we detail the
		 baseline reconfigurable GPU explored in this paper.

		 \begin{figure}[t!] \centering
\includegraphics[width=0.485\textwidth]{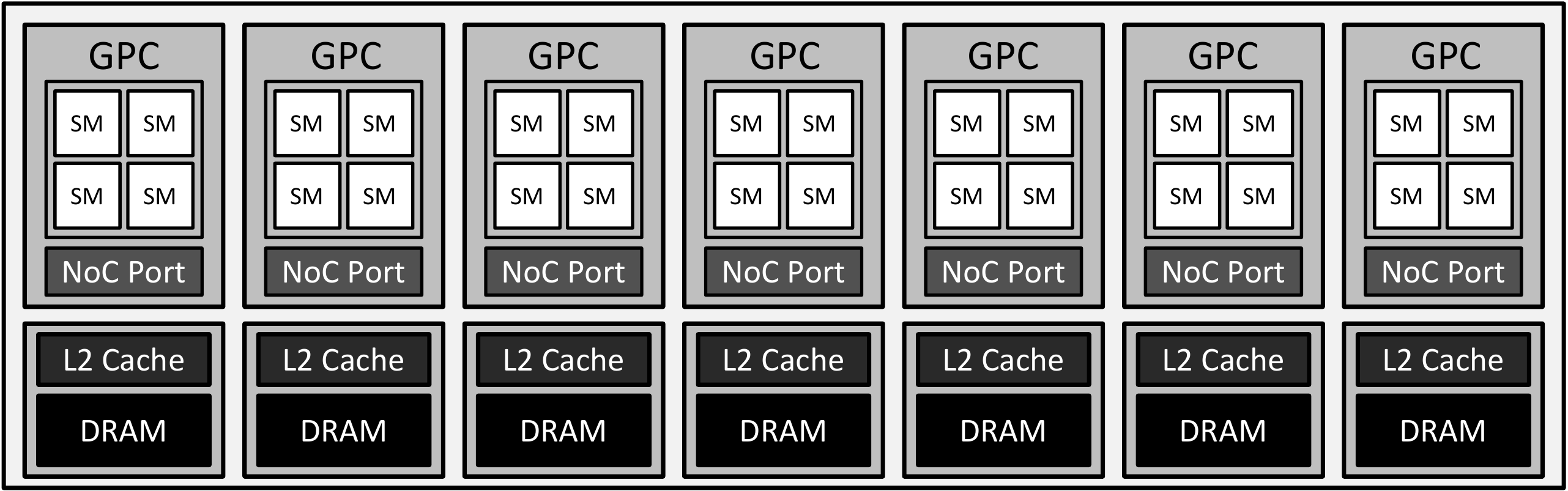}
\caption{
High-level overview of an NVIDIA GPU architecture.
}
\label{fig:gpu_arch}
\end{figure}

\subsection{A ``Reconfigurable'' GPU Architecture}
\label{sect:gpu_arch}

As this paper utilizes NVIDIA's MIG-enabled GPU as
a vehicle to construct a reconfigurable multi-GPU inference server, we use
NVIDIA's A100 GPU~\cite{a100} to describe a modern SIMT (single-instruction multiple-thread) based
GPU architecture. In the remainder of this paper, we use terminologies defined in NVIDIA's CUDA programming language~\cite{cuda}.

{\bf GPU hardware architecture.}
\fig{fig:gpu_arch} provides an overview of our baseline GPU architecture. 
The most fundamental computational building block of a GPU is an SM (streaming multiprocessor),
which is a SIMD vector processor (but programmed using the SIMT programming semantics
which is different than traditional vector programming). Each SM contains a large register-file to enable GPUs
to employ massive number of threads to concurrently execute with fine-grained, hardware-level
context switching for latency hiding. An SM also contains an L1 cache and scratchpad memory that can capture
high-locality datasets within the vicinity of our SIMD processor. Multiple SMs are grouped into a cluster, which is called
a GPC (Graphics Processing Cluster) and the SMs within the same GPC share the communication ports to the NoC (network-on-chip). As GPUs are throughput-optimized processors, the NoC is implemented using a high-bandwidth crossbar. The crossbar that interconnects multiple GPCs
are utilized to access the L2 cache/DRAM slices, which allows an L2 cache miss to be routed to the corresponding off-chip memory channel to access DRAM.

{\bf GPU software architecture.} CUDA employs the SPMD (single-program
		multiple-data) programming model, where a single program (the
			\emph{kernel}) gets executed by \emph{all} the threads that are spawned
		for execution. The programmer is expected to group the threads into a
		granularity called \emph{thread-blocks} or \emph{concurrent thread-arrays}
		(aka CTAs) and the hardware-level scheduler is in charge of scheduling
		CTAs to the SMs for execution. Once a CTA is scheduled to a given SM, it stays there until the entire program's execution
		is finalized (i.e., a scheduled CTA does not migrate to other SMs). 

\begin{figure}[t!] \centering
\includegraphics[width=0.37\textwidth]{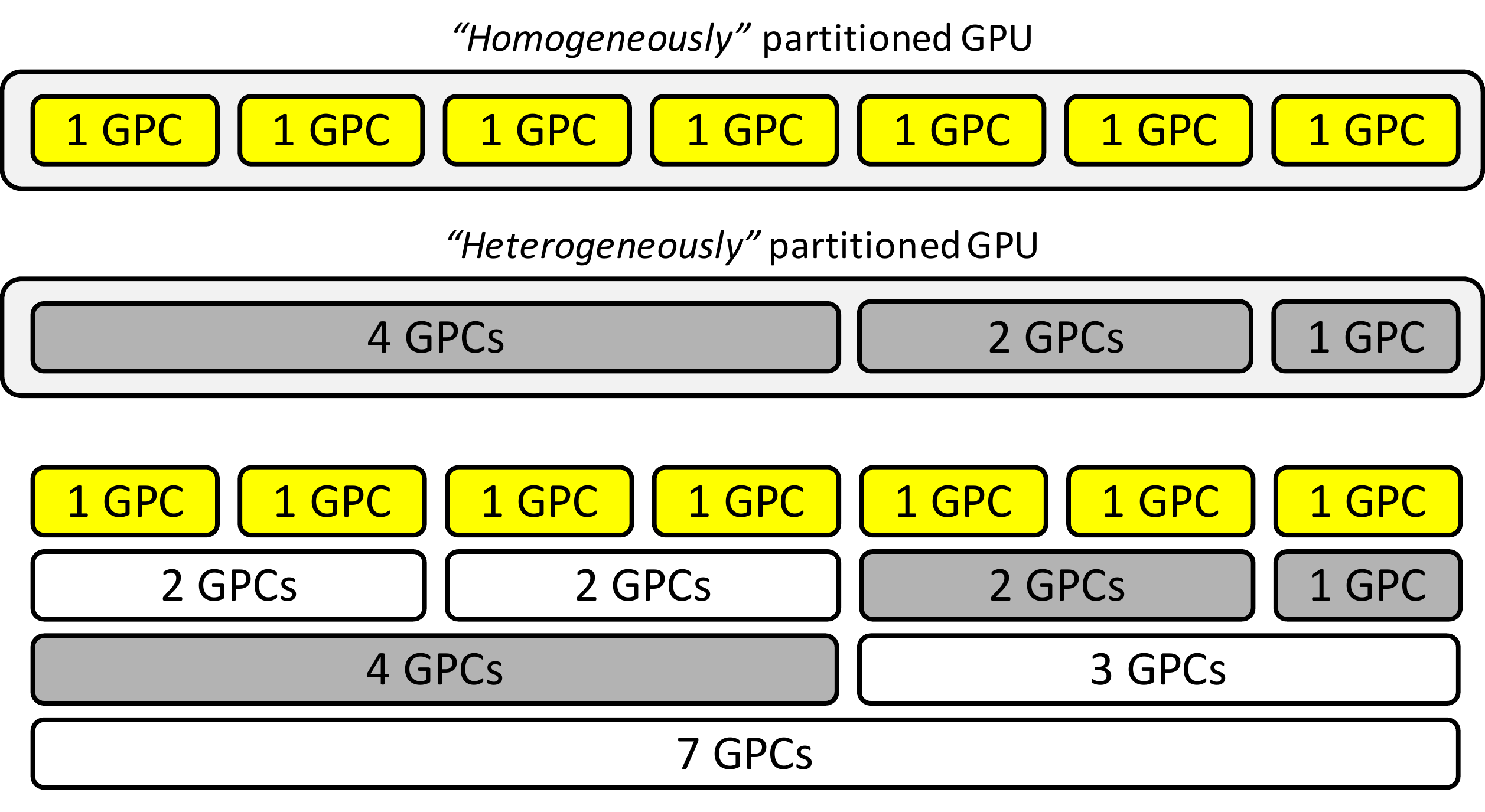}
\caption{
Example configuration of GPU partitions.
}
\label{fig:gpu_mig}
\end{figure}

{\bf Adding reconfigurability to the GPU.} In A100, 
	the GPCs (compute) and the L2/DRAM slices (memory) are utilized as basic building
	blocks to architect a GPU with reconfigurability. 
	Specifically, a  GPU \emph{partition}
	can be defined at the granularity of a GPC, so A100 which contains seven GPCs can be configured
	up to seven GPU partitions (each partition having just a single GPC worth of compute capability).
	\fig{fig:gpu_mig} illustrates valid GPU partition combinations available in A100, allowing it to be
	(re)configured into one big GPU ($7$ GPCs) or multiple small ($1$ or $2$ GPCs) or medium ($3$ or $4$ GPCs) sized GPUs. The reconfigurable GPU is provided with the proper architectural support for SR-IOV (single root input/output virtualization), so each GPU partition is given the necessary hardware-level features to function as a true ``standalone'' GPU device, i.e., each GPU partition can be handed over to a process or a VM, with performance isolation guarantees.

\subsection{Related Work}
\label{sect:related}

Utilizing multi-GPU systems for ML inference and training has been studied
extensively in prior literature.  DjiNN and
Tonic~\cite{djinn_and_tonic} is one of the early works on ML inference servers
based on a homogeneous set of GPU devices, presenting an open-source software
infrastructure for deploying ML services at datacenters. Recent ML frameworks
like TensorFlow Serving~\cite{tf_serving}, AWS SageMaker~\cite{amazon_sagemaker}, and NVIDIA Triton
Inference Server~\cite{trtis} are also dedicated software packages intended to ease
the development of ML inference servers. In terms of ML training,
		PipeDream~\cite{pipedream}, GPipe~\cite{gpipe}, and Megatron-LM~\cite{megatron-lm} (among many others)
	utilize multi-GPU systems for training large-scale ML models. None of these prior studies
	utilize the reconfigurable GPU we explore in this paper, rendering the 
	key contributions of our work stand on its own.

In terms of leveraging the idea of
\emph{heterogeneous} computing for ML inference, DeepRecSys~\cite{deeprecsys} employs a heterogeneous
CPU-GPU system for servicing recommendation services. MOSAIC~\cite{mosaic}, uLayer~\cite{mulayer},
	and JointDNN~\cite{jointdnn} explore the possibility of utilizing the heterogeneous compute
	capabilities within mobile devices (e.g., CPU, GPU, NPUs, DSPs) for accelerating ML inference.
These prior art primarily focus on partitioning the DNN model and scheduling them across the heterogeneous
processing units. Our work on the other hand focuses on the partitioning of the reconfigurable GPU
hardware rather than the ML model. Overall, the key contribution of this paper is
orthogonal to these prior studies.

\section{Characterization and Motivation}
\label{sect:motivation}

 To the best of our knowledge, this paper is the first to conduct a detailed
 characterization on the utility of reconfigurable GPUs for ML inference
 servers. Given there are virtually no prior work that explores this research
 space, we assume the following designs as the baseline starting point for
 reconfiguration, i.e., partitioning the monolithic GPU into a
 \emph{homogeneous} set of small, medium sized GPU partitions or using it as
 one large GPU as-is. The rest of this paper refers to a GPU partition
 configured with a) one or two GPCs as \Small{1}/\Small{2} b) three or four
 GPCs as \Medium{3}/\Medium{4}, and c) all seven GPCs as \Largest{7}. 
 In terms
 of benchmarks, we study three DNN models with different levels of
 compute-intensity, MobileNet (low), ResNet (medium), and BERT (high).
	All the results presented in this section focus on a \emph{single} instance
	of a particular GPU partition as means to characterize the different partition
	granularities' unique computation power and its GPU utilization vs. latency tradeoff
	properties. Later in \sect{sect:eval}, we evaluate our proposal over a multi-GPU
	server equipped with $8$ A100 GPUs. 
\sect{sect:methodology} details our evaluation methodology further.

\begin{figure}[t!] \centering
\includegraphics[width=0.45\textwidth]{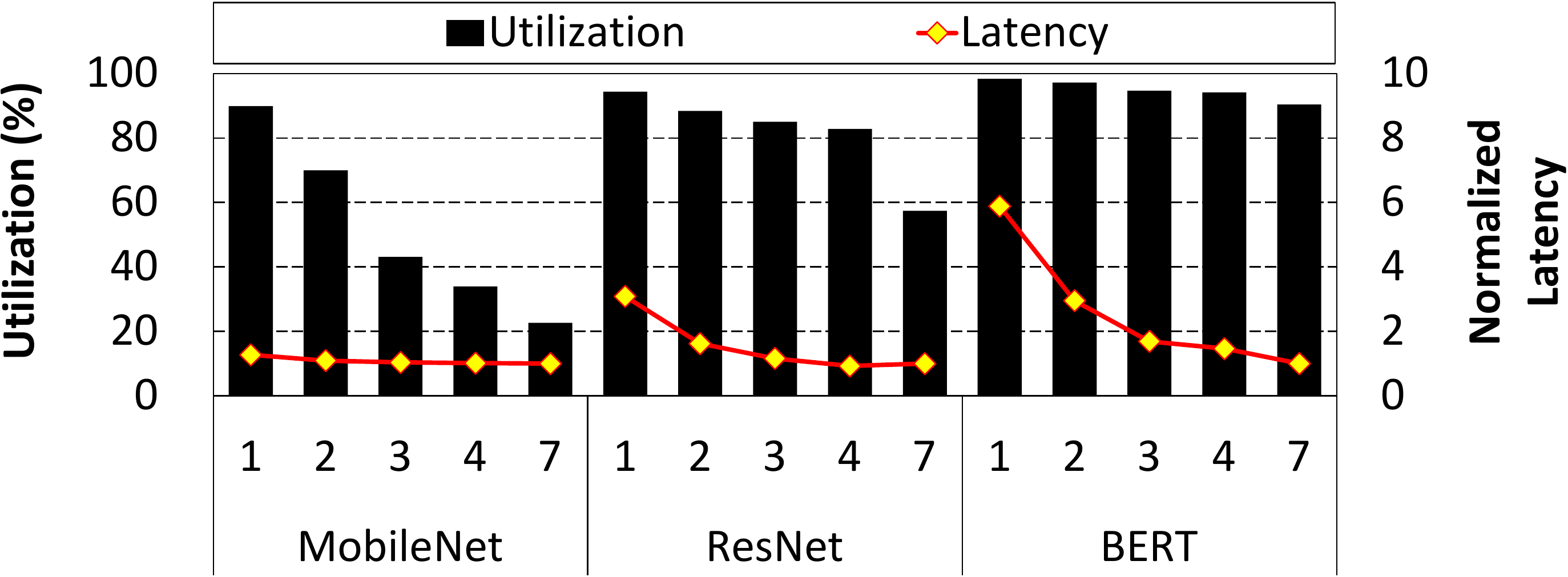}
\caption{
Effect of the GPU paritition size (x-axis, from \Small{1} to \Largest{7}) 
	on GPU compute utilization (left-axis) and latency (right-axis). Experiment
	assumes a batch size of $8$ executed over a single GPU partition.
}
\label{fig:motivation_util_vs_latency}
\end{figure}

\begin{figure*}[t!] 
\centering
\subfloat[]{
\includegraphics[width=0.91\textwidth]{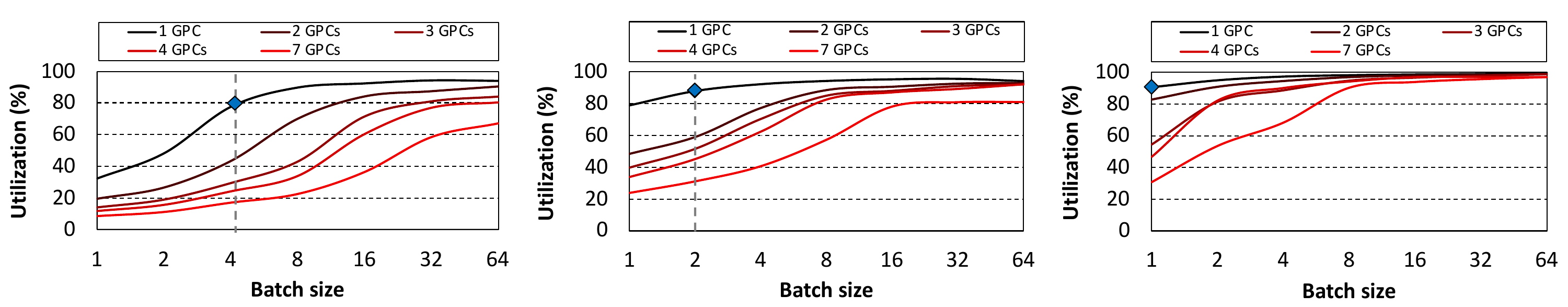}
\vspace{-1.5em}
	\label{fig:}
}
\vspace{-1em}
\subfloat[]{
	\includegraphics[width=0.91\textwidth]{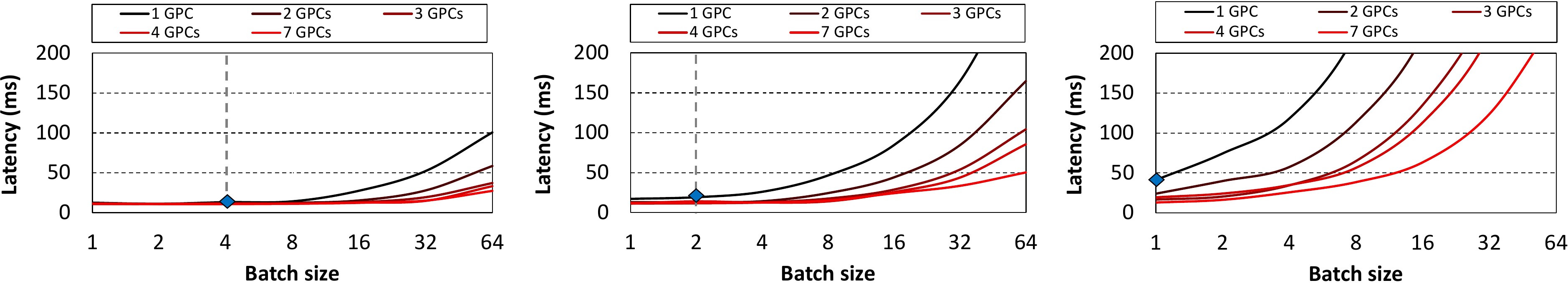}
	\vspace{-1.5em}
	\label{fig:}
}
\caption{ 
Effect of batch size on reconfigured GPU's (a) GPU utilization and (b) average latency for MobileNet (left), ResNet (middle), and BERT (right). The blue diamonds refer to the \knee points of \Small{1} for each of the DNN models.
}   
\label{fig:motivation_batch_effect}
\end{figure*}

\subsection{Effect of Model Size on Latency \& Server Utility}
\label{sect:partition_size_effect}

\fig{fig:motivation_util_vs_latency} shows the compute utilization and latency
of the reconfigurable GPU when we sweep the size of each partition from the
smallest \Small{1} to largest \Largest{7}.  Under small partition sizes like
\Small{1}, all DNN models universally achieve high GPU utilization.  As such, a
simple yet intuitive partitioning strategy would be to statically partition the
reconfigurable GPU into a homogeneous set of small GPUs (i.e., partition into
		seven \Small{1}), addressing the GPU underutilization problem in hand.
However, blindly partitioning the large GPU into smaller ones without
considering the unique computation demands of the target model can be
suboptimal.  This is because the reduced compute capability of a small GPU may
not be sufficiently high enough for the DNN model, leading to significantly longer
latency and violating SLA.  For instance, while both
MobileNet and ResNet are DNN models for computer vision applications, the
computation requirements			of MobileNet are much more lightweight than
ResNet as MobileNet heavily employs compute-efficient $1\times1$ convolutions
as well as depthwise filters. Consequently,  ResNet  experiences a more steep
increase in latency when the GPU partition size is decreased because it's
performance becomes more sensitive to the (relatively) smaller computation
power of \Small{1,2} than the lightweight MobileNet. The same principle
holds for the compute-intensive BERT, exhibiting the highest increase in
latency when smaller GPU partition sizes are employed.

Overall, we conclude that determining an optimal partitioning granularity for
reconfigurable GPUs requires careful consideration of each model's unique
algorithmic properties and its compute/memory demands. For instance, our
experiment in \fig{fig:motivation_util_vs_latency} shows that the optimal 
partitioning point
for ResNet is around \Medium{3} as it does not incur significant
increase in latency while the achieved GPU utilization is reasonably high. The sweet
spot for MobileNet on the other hand is \Small{1} as it achieves
approximately $2\times$ higher GPU utility  while ``only'' experiencing a
latency increase of $23\%$ vs. \Medium{3}.  In general, our
characterization demonstrates the challenges and pitfalls of a
``one-size-fits-all'' approach, i.e., partitioning the reconfigurable GPU into
a homogeneous set of GPU partitions, as no single partitioning granularity
could universally fulfill the various DNN model's computation demands as well as its individual latency goals.

\subsection{Effect of Batch Size on Latency \& Server Utility}
\label{sect:batch_size_effect}

Along with the individual DNN's model specific properties, the batch size of a
query is another key factor that affects GPU utilization and latency, posing
another challenge in finding the optimal partitioning granularity.  Inference queries
with large batch sizes help increase GPU utilization as it better exploits
parallelism and locality across the batched inputs. On the other hand, large
batches increase the amount of computations so it can adversely
affect the level of SLA violations when the latency is increased to an
unacceptable level.

\fig{fig:motivation_batch_effect} shows the effect of batch size on our
reconfigured GPU's compute utilization and average latency.  In general, all
models over all GPU partition sizes experience a monotonically increasing GPU
utilization and latency as the batch size is increased. However, once the GPU
utilization reaches a plateau around $80-90\%$, the latency increases much more rapidly
with larger batch sizes.  This is because executing with a larger batch size only
helps improve GPU utilization incrementally when the utility already neared its
peak value, while the proportionally increased computation directly translates
into a linear increase in execution time. 
We hereafter refer to this point
as the ``\emph{max
batch size at the knee of the latency curve}'', or \knee
in short (e.g., denoted as blue diamond shapes for \Small{1} in \fig{fig:motivation_batch_effect}). Naturally, the \knee differs significantly across
different GPU partition sizes or DNN model types, with small GPU partitions generally having a smaller \knee while larger GPU partitions having a larger \knee.

Overall, large models like BERT
are able to achieve high GPU utilization under small GPU partitions even when
the batch size is small.  Therefore,  executing large batches of BERT on a small
\Small{1} is likely to be a poor scheduling decision as the benefits in GPU
utility is minimal while the penalty in latency is high.  \Small{1} however is
a reasonable design point for the lightweight MobileNet as it does a  much
better job in handling medium-to-large batches, achieving high GPU utility
while minimally sacrificing latency. 

Given such, one might choose to utilize the results in \fig{fig:motivation_batch_effect} to
\emph{manually} determine a model specific and batch size specific partitioning
point that balances GPU utilization and latency. Unfortunately, the size of an
input batch can vary significantly per inference server's average
query size distribution (i.e., a log-normal distribution for datacenter web-services, \sect{sect:training_vs_inference}). As a result, a ``one-size-fits-all'',
	homogeneous partitioning strategy (even if it is hand-tuned on a per-model
			basis) again is not able to robustly capture the various query sizes (i.e.,
				batch sizes) routed to the inference servers.

\subsection{Our Goal: A Heterogeneously Partitioned GPU Inference Server and Its Scheduling Algorithm}
\label{sect:our_goal}

{\bf A ``heterogeneous'' multi-GPU ML inference server.}
Overall, our characterization revealed two key challenges with a homogeneously
partitioned multi-GPU inference server. First, a statically chosen, fixed
partitioning granularity is not able to efficiently capture the
\emph{model specific} computation diversity of DNNs, failing to achieve
 low latency and high GPU utilization simultaneously.
Second, the dynamically varying input batch size poses another problem because
a rigidly configured, single-granular GPU partition size cannot flexibly adapt
to the varying computation requirements of input batches.    Rather than having multiple, identical
instances of a single GPU partition size (e.g., six instances of
		\Small{1} or three instances of \Medium{2}), our proposed
{\bf P}artitioning {\bf A}lgorithm for {\bf R}econfigurable multi-GPU {\bf I}nference {\bf S}ervers
(\paris) 
partitions the reconfigurable GPUs into a
\emph{heterogeneous} set of GPU partitions.
As we detail in the next section, 
\paris
	systematically evaluates both the target model's inference properties
\emph{and} the input query size distribution to derive a fruitful set of
multi-granular partitioning sizes as well as the number of instances to
deploy for each partition size. The
collection of GPU partitions with heterogeneous compute capabilities enable our
proposed ML inference server to flexibly respond and adapt to the
model specific compute demands of DNNs as well as the dynamically changing
query sizes.

\begin{figure}[t!] 
\centering
\subfloat[]{
\includegraphics[width=0.4\textwidth]{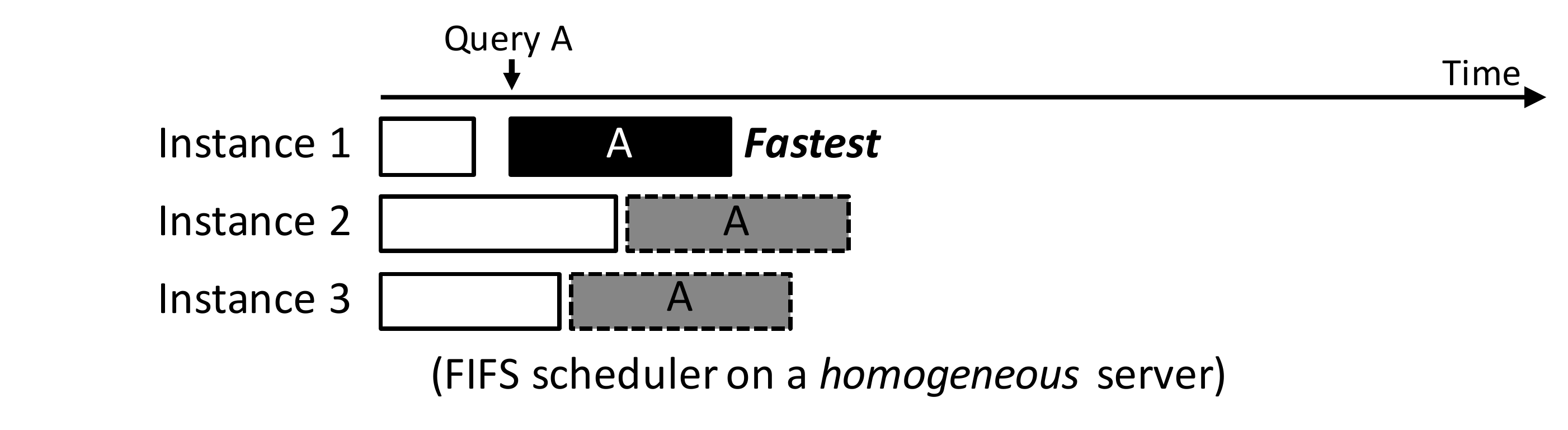}
\vspace{-1.5em}
	\label{fig:}
}
\vspace{-1em}
\subfloat[]{
	\includegraphics[width=0.4\textwidth]{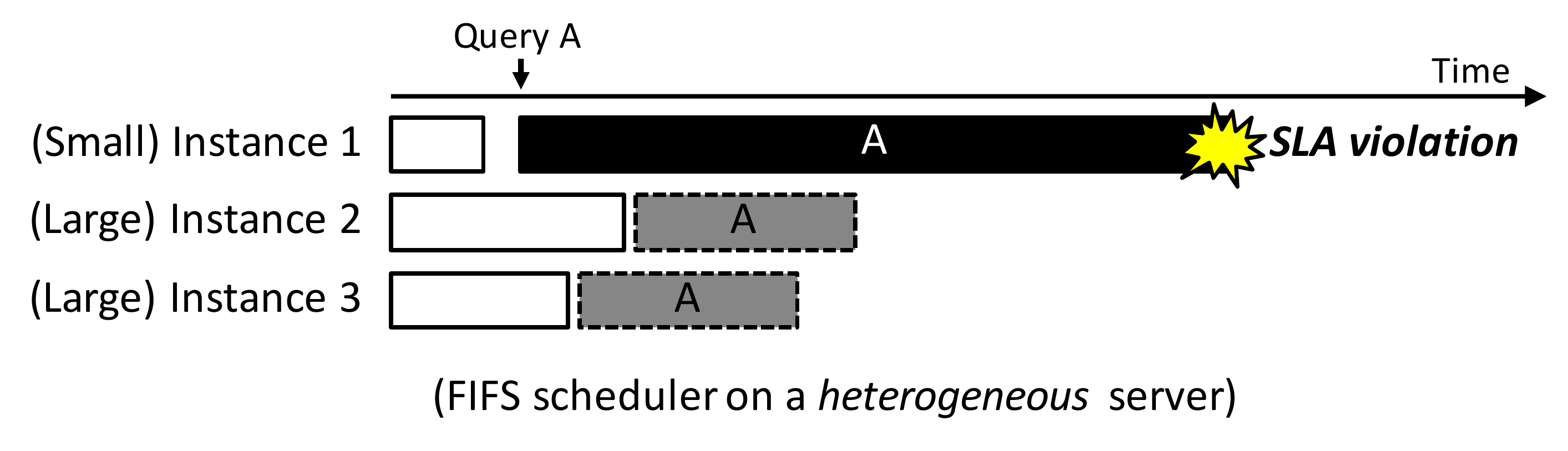}
	\vspace{-1.5em}
	\label{fig:}
}
\caption{ 
Timeline of FIFS policy when adopted on a (a) homogeneously and (b) heterogeneously partitioned multi-GPU inference server.
}   
\label{fig:motivation_fifs_effect}
\end{figure}

{\bf A ``heterogeneity-aware'' scheduling algorithm.} As \paris enables the inference server's compute capability to become diverse, a scheduling algorithm that best exploits such
heterogeneity is in need.  Current state-of-the-art multi-GPU inference servers
(e.g., NVIDIA Triton Inference Server~\cite{trtis}) employ a first-idle
first-serve (FIFS) scheduling policy where the newly inserted inference query
is scheduled to an \emph{idle} GPU available in the system. As depicted in
\fig{fig:motivation_fifs_effect}(a), an FIFS scheduling policy is both intuitive
and cost-effective for 
homogeneous multi-GPU system to
minimize the number of idle GPUs and reduce average latency. Under our proposed,
heterogeneous multi-GPU system however, FIFS can lead to suboptimal scheduling
decisions as it fails to accommodate the diverse computation power of our GPUs.
In \fig{fig:motivation_fifs_effect}(b), we assume a heterogeneously partitioned
multi-GPU server with two large and one small GPU. When query $A$ arrives to
the server, the FIFS scheduler chooses the small GPU for execution as it is the
only  idle GPU available. Because the idle GPU is a small one,
			the latency to service this query is \emph{longer} than what would have
			been experienced had the idle GPU been a large one, leading to an SLA violation.
			Consequently, a
			better scheduling decision would have been to wait until any one of the
			large GPUs complete its current query and schedule query $A$ there
			instead. The baseline FIFS however is unaware of the heterogeneous
			computing power in our \paris server, leading to longer latency and aggravating overall
			performance.  We propose an {\bf EL}astic {\bf S}cheduling {\bf
				A}lgorithm (\elsa) that is designed with \emph{heterogeneity-awareness} in mind,
				maximally exploiting the potential of the
				heterogeneous computing power of our \paris multi-GPU system. 
We now detail our two proposals, \paris and \elsa.

\section{Proposed Architecture: PARIS and ELSA}
\label{sect:proposed}

\subsection{High-level Overview}
\label{sect:overview}

\begin{figure}[t!] \centering
\includegraphics[width=0.485\textwidth]{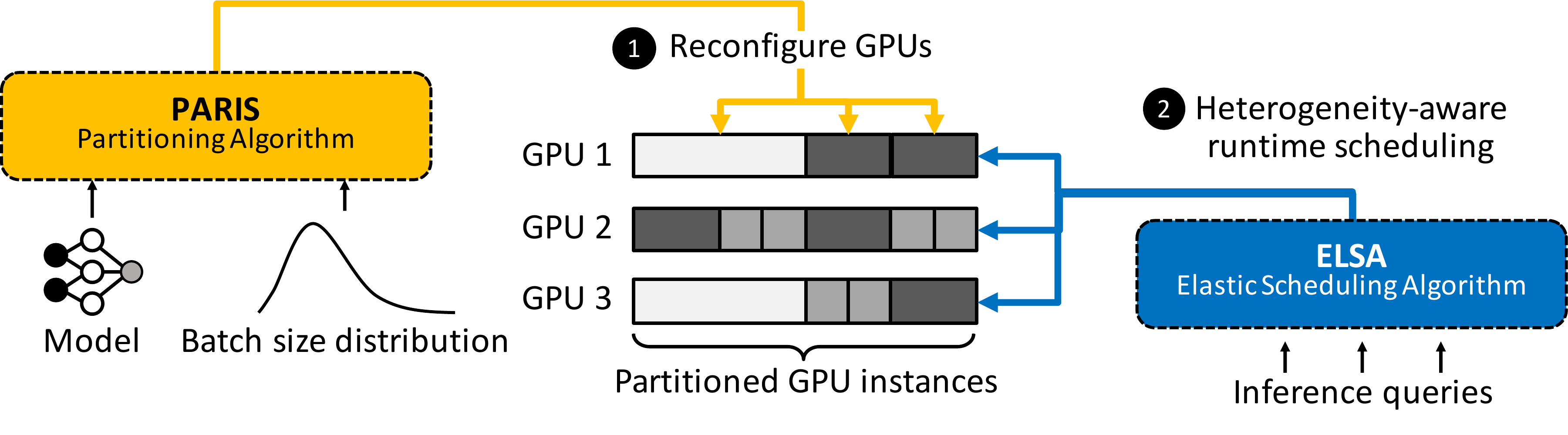}
\caption{
High-level overview of our proposed ML inference server.
}
\label{fig:proposed_server}
\end{figure}

\fig{fig:proposed_server} provides an overview of a ML inference server
employing our two proposals, \paris and \elsa.  In this section, we first make
a case for partitioning the reconfigurable GPUs \emph{heterogeneously} 
using \paris (\sect{sect:paris}). \paris utilizes both the
model specific inference properties (e.g., latency vs. GPU utility under a
		target GPU partition size) and the batch size distribution information to
systematically generate a heterogeneous set of partitioning granularities as
well as the number of instances to deploy for each partition.  Our second
proposition \elsa is a high-performance scheduling algorithm co-designed with
our heterogeneous \paris inference server (\sect{sect:elsa}).  \elsa uses a
heterogeneity-aware, inference latency prediction model to estimate a given
query's SLA slack and determine which among our heterogeneous GPUs are best
suited to service the query. As we detail in this section, \elsa's
heterogeneity-awareness  helps maximize server utilization while minimizing SLA
violations.

\subsection{PARIS}
\label{sect:paris}

We first discuss the key insights that motivate our \paris, followed by discussions
on its design and implementation.

{\bf Key observations.} Our characterization in \sect{sect:batch_size_effect}
revealed that the \emph{max batch size at the knee} (\knee) 
	varies significantly across different GPU partition sizes, with smaller (larger) GPU
	partitions having smaller (larger) \knee. Based on this
	characterization study, we make several key observations that motivate \paris
	as follows:

	\begin{enumerate}
	\item For any given GPU partition size, having it handle batch sizes larger
	than its \knee is not cost-effective as the gains in
	GPU utilization is minimal while the penalties in latency can be significant.

	\item Assuming the input batch size to execute is smaller than the \knee
	for a given model, small (medium) GPU partitions are generally more
	cost-effective when handling small (medium) batch sizes than large GPU partitions
	as it can achieve high GPU utility while not sacrificing latency.

	\item Similarly, large GPU partitions are efficient when handling large 
	batch sizes as it does not incur too high of a latency overhead (thanks to its high
			computation power) while still achieving high GPU utilization. 
 While scheduling small batches (smaller than the \knee) 
	to  large GPU partitions is certainly feasible, it can suffer from low 
	GPU utilization. Consequently, small(er) batches are best when 
	delegated to small(er) GPU partitions rather than scheduling them to 
	large(r) GPUs.
	\end{enumerate}

\begin{figure}[t!] 
\centering
\subfloat[]{
\includegraphics[width=0.4\textwidth]{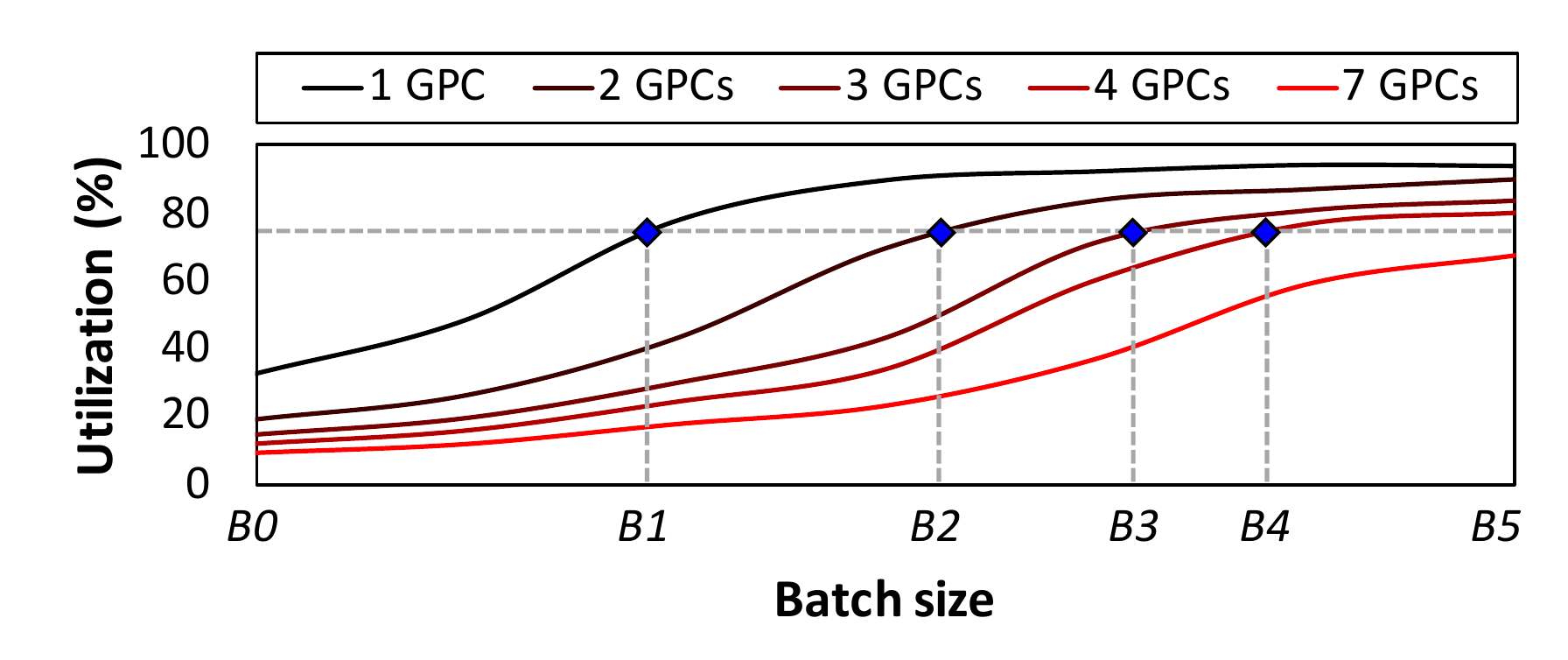}
	\label{fig:}
}
\vspace{-1em}
\subfloat[]{
\includegraphics[width=0.4\textwidth]{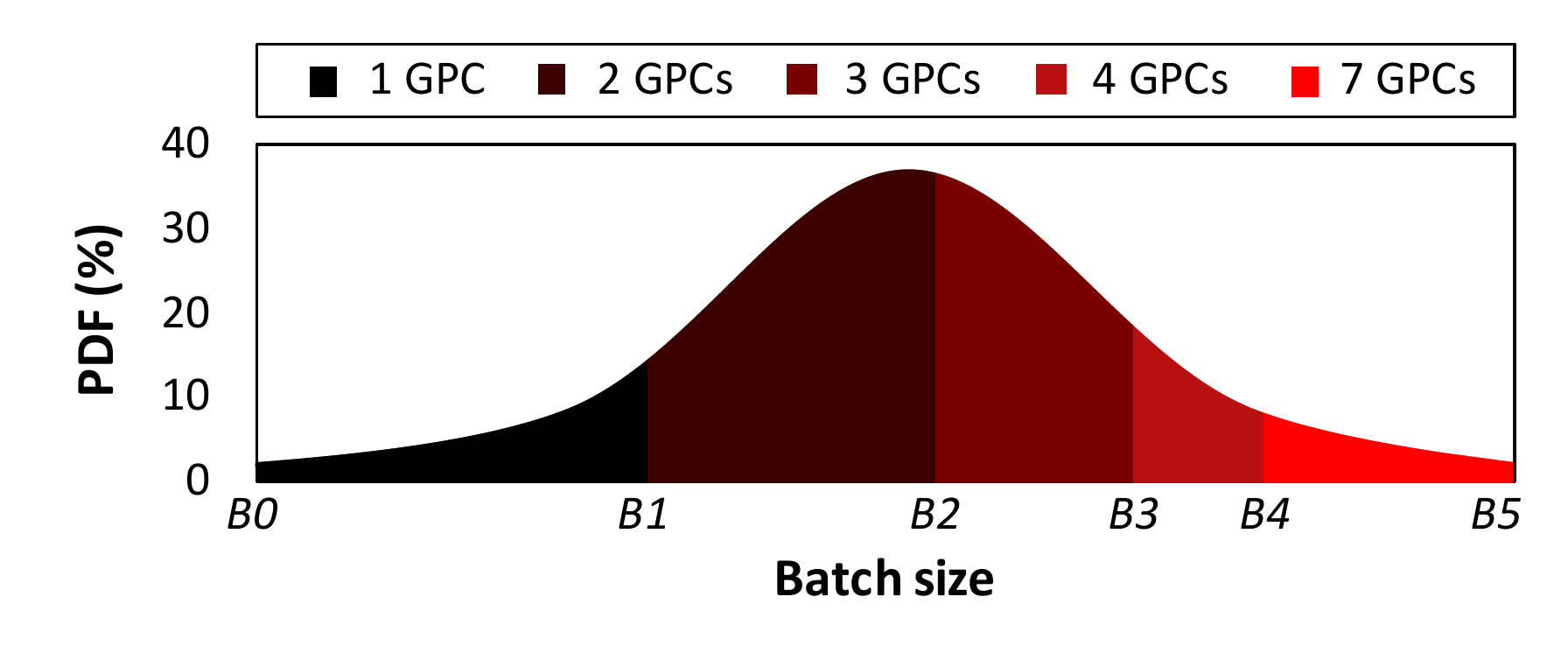}
	\label{fig:}
}
\caption{ 
Considering both (a) model specific inference properties and (b)
	batch size distribution simultaneously for GPU reconfiguration in \paris.
}   
\label{fig:proposed_paris_motivation}
\end{figure}

{\bf Partitioning with both model specific properties ``and'' batch size distribution in
	mind.}  \fig{fig:proposed_paris_motivation} visualizes our key approach that
	incorporates both the model specific latency properties as well as the varying
	input batch sizes as part of our partitioning algorithm. We first conduct a
	one-time profiling of the [GPU utilization vs. latency] curve per each GPU
	partition size, which was also used in our characterization in
			\fig{fig:motivation_batch_effect}. Using the characterization results,
	we are able to derive each GPU partition's \knee for a
	target DNN model (B1/B2/B3/B4/B5 for the five GPU partition sizes in
			\fig{fig:proposed_paris_motivation}(a)).  These \knee values	are then utilized to \emph{split} the batch size distribution
into multiple, non-overlapping segments of batch size ranges ([B0-B1], [B1-B2], $\ldots$), where
the number of segments matches the number of GPU partitioning granularities we consider in
\paris (\fig{fig:proposed_paris_motivation}(b)).  The batch size distribution is virtually a probability density
function (PDF) that models the likelihood of a particular batch size to be
queried to the inference server, one which is known to follow a log-normal
distribution in web-services (\sect{sect:training_vs_inference}). This function
can readily be generated in the inference server by collecting the number
of input batch sizes serviced within a given period of time, which \paris can
utilize as a proxy for the batch size distribution PDF.
Each of the partitioned batch range segments are then assigned to
its dedicated GPU partitions one-by-one, the $n$-th smallest batch range segment assigned
to the $n$-th smallest GPU partition (\fig{fig:proposed_paris_motivation}(b)).

The key benefits of our partitioning mechanism is clear. Because the profiled,
		per-model characterization curves (\fig{fig:motivation_batch_effect}) are
		used to derive the \knee values, \paris can accommodate the model specific
		utilization-vs-latency tradeoff properties into our partitioning algorithm.
		Additionally, each GPU partition now has a dedicated batch range
		segment to service that best suits its compute capabilities (which is governed by the
				batch size distribution \emph{and} the \knee values), so \paris can 
		better handle
		the diverse query sizes routed to the inference server with high
		utilization using its heterogeneous GPU partitions.

		\begin{figure}[t!] \centering
\includegraphics[width=0.46\textwidth]{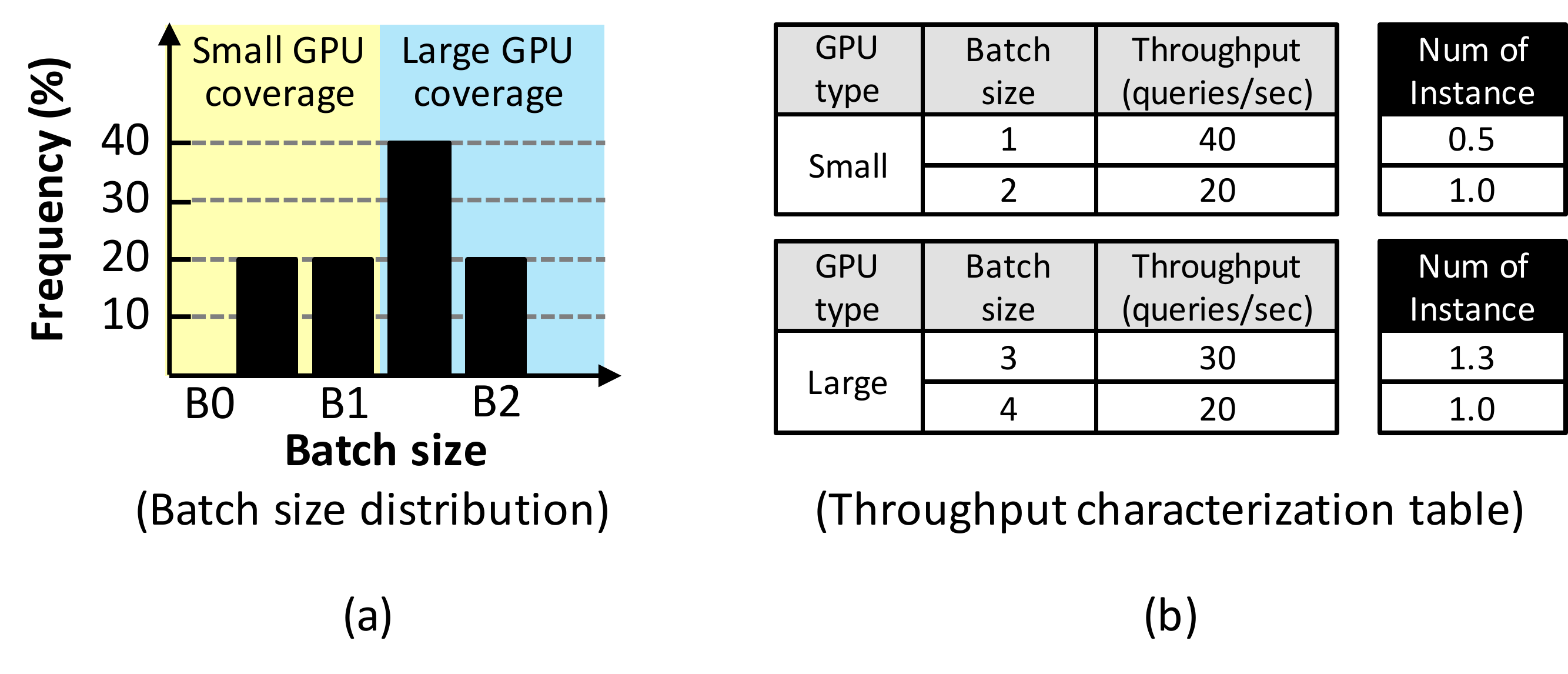}
\caption{
Example showing how \paris derives the number of instances per each GPU partition.
}
\label{fig:proposed_paris_example}
\end{figure}

{\bf Determining the number of partition ``instances''.} As \paris has now
determined which batch size range the partitioned GPUs will be handling, a
derivation of \emph{how many instances} of these GPU partitions should be
deployed is required.  Two factors must be considered in determining the
optimal number of instances to deploy: 1) the likelihood of a particular batch
size to be queried to the inference server (which is reflected in the batch size distribution PDF), and 2) the
\emph{effective} inference throughput of a particular GPU partition when
handling its responsible batch range segment (which is derived using our
		profiled characterization graph in \fig{fig:motivation_batch_effect},
		i.e., number of queries serviced/second).  We use
		\fig{fig:proposed_paris_example} as a driving example to 
		explain our mechanism that derives the number of instances required per each partition
		size. We assume that up to two GPU partition sizes are available, each of which has a
		\knee value of B1(=$2$) and B2(=$4$), respectively. Therefore, the small
		(large) GPU covers batch size $1$/$2$ ($3$/$4$), which accounts for
		$20$+$20$=$40\%$ ($40$+$20$=$60\%$) of the inference query traffic as estimated
		through the batch size distribution PDF
		(\fig{fig:proposed_paris_example}(a)). Consider the small GPU which is
		measured and estimated (through profiling) to provide an effective inference
		throughput of $40$ and $20$ queries/sec for batch size $1$ and $2$, respectively (\fig{fig:proposed_paris_example}(b)). Assuming
		the total number of queries the inference server needs to service is $100$,
		we can expect $20$ queries of batch size $1$, $2$, and $4$ each, and $40$ queries of
		batch size $3$.
	Now, because the
		effective throughput for batch size $1$ is two times higher than that for
		batch size $2$ ($40$ vs. $20$ queries/sec), we virtually need $0.5$ (=$20$/$40$, i.e., number of queries to be serviced for a given batch/effective throughput for that batch) small GPU to sufficiently serve batch $1$ queries
		and another $1$ (=$20$/$20$) small GPU to service batch $2$ queries, requiring $1.5$ (=$0.5$+$1.0$) small GPUs in aggregate.
Similarly, a total of $2.3$ large GPUs is in need to fully service batch $3$/$4$ (\fig{fig:proposed_paris_example}(b)).
The ratio of (1.5:2.3)=(number of small GPU instances:number of large GPU instances) can therefore be
utilized to determine by what fraction should \paris divide up the available compute resources
within our multi-GPU server (i.e., total number of GPCs per GPU $\times$ number of GPUs per server).
Below we detail the implementation aspects of \paris.

\begin{algorithm}[t!]
\caption{PARIS}
\label{algo:paris}
\begin{algorithmic}[1]
\scriptsize
\rmfamily
\Procedure {Partitioning\_Algorithm()}{}
\State $GPC[k] = [Possible \ configurations \ of \ GPU \ partition \ size]$
\State $Dist[b_{1}, b_{2}, ..., b_{n}] = [p_{1}, p_{2}, ..., p_{n}] \ (0 \leq p_{n} \leq 1)$
\State $Util_{k} [b_{1}, b_{2}, ..., b_{n}] = [u_{1}, u_{2}, ..., u_{n}]_{k} \ (0 \leq u_{n} \leq 1)$
\State $Throughput_{k,b} =  Throughput \ of \ GPU \ configuration \ k \ in \newline \hspace*{10.7em} batch \ size \ b$
\Statex
\Statex \LineComment{Step A: Find \knee under each GPU partition (one-time cost)}
\For {$k = 1 \ \textbf{to} \ size(GPC)$}
\State $Find \ B_{k} \ that \ Util_{k}[B_{k}] \geq 0.8 $
\EndFor
\Statex
\Statex \LineComment{Step B: Derive the relative ratio of GPU partition instance numbers}
\For {$k = 1 \ \textbf{to} \ size(GPC)$}
\State $R_{k} \leftarrow 0$
\For {$b = B_{k-1}+1 \ \textbf{to} \ B_{k}$}
\State $R_{k} \leftarrow R_{k} + {\frac{Dist(b)}{Throughput_{k,b}}}$
\EndFor
\EndFor
\Statex
\Statex \LineComment{Step C: Determine the absolute number of GPU partition instances}
\State $sum_R \leftarrow 0$
\For {$k = 1 \ \textbf{to} \ size(GPC)$}
\State $sum_R \leftarrow sum_R + (GPC[k] \times R_k)$
\EndFor
\State $C \leftarrow {\frac{Total \ number \ of \ available \ GPCs}{sum_R}}$
\For {$k = 1 \ \textbf{to} \ size(GPC)$}
\State $\mathrm{N_{k} \leftarrow C \times R_{k} }$
\EndFor
\State \Return $[N_{1}, N_{2}, ..., N_{k}]$
\EndProcedure
\end{algorithmic}
\end{algorithm}

{\bf Implementation.} \algo{algo:paris} is a pseudo-code of \paris, putting all
of the pieces discussed in this subsection together. The three most important
input data to \paris is 1) the PDF of batch size distribution (\emph{Dist[]},
	line $3$), 2) a GPU partition's compute utilization at a particular batch
	size (\emph{Util[]}, line $4$), and 3) the effective inference throughput of
	a particular GPU partition when executing a particular batch size
	(\emph{Throughput$_{k,b}$}, line $5$).  Assuming there are $k$ possible GPU
	partition sizes available within the reconfigurable GPU (\emph{GPC[]}, line
			$2$), \paris first initiates a one-time derivation of the \knee for each of the
	$k$ partition sizes using the profiled [GPU utilization-vs-latency] curve (line $6$-$9$).
	For clarity of explanation, we assume the batch size that a given GPU
	partition starts exceeding $80\%$ GPU utilization is the \knee value, which is stored into \emph{B$_{k}$} (line $8$). Once \emph{B$_{k}$}
	is derived, \paris uses the set of \knee values to determine the
	ratio between each GPU partition's required number of instances (line $10$-$16$),
	as explained through the example in \fig{fig:proposed_paris_example}. Finally,
	the derived relative ratio is used to determine the absolute number of instances 
	a particular GPU partition size should be instantiated with (line $17$-$26$), which
	is utilized to configure our \paris-enabled heterogeneous multi-GPU server.

\subsection{ELSA}
\label{sect:elsa}

Once \paris is applied to our reconfigurable multi-GPU system, the scheduler is
given a selection of heterogeneous computing devices it must judiciously
utilize for maximum efficiency. As discussed in \sect{sect:our_goal}
(\fig{fig:motivation_fifs_effect}), the baseline FIFS scheduling algorithm
fails to accommodate the diverse compute capabilities of our heterogeneously
partitioned \paris system, leading to aggravated latency and GPU utility. 

Our \elsa is designed with ``heterogeneity-awareness'' in mind and consists of
three major components:

\begin{enumerate}
\item First, 
we propose a profiling-based approach in \emph{estimating} a DNN model inference query's
execution time when scheduled to a particular GPU 
partition. 

\item The estimated DNN execution time is then used to calculate the remaining
SLA \emph{slack time} for that query.

\item Finally, the  SLA slack time
is utilized by our scheduler to dynamically judge 
which among
the heterogeneous GPU partitions would this query be best served by, with
minimizing SLA violations as a topmost scheduling objective. 
\end{enumerate}

We now detail each of these three components below.

{\bf Estimating DNN model execution time via profiling.} A key observation of
our profile-based approach is that a DNN model's inference execution time
over a target GPU architecture is highly deterministic and predictable.  Prior
work~\cite{gao:2017:tetris,nexus,grandslam,prema,lazybatching} similarly
observed the deterministic nature of DNN inference latency and \elsa's DNN
model execution time estimator leverages such property for slack estimation. Specifically, we
conduct an exhaustive, one-time profiling of a target DNN model's execution
time over a target GPU partition size and all possible batch sizes. The latency
to collect this information for all the design points we consider is
approximately $5$ minutes, which is a one-time cost and is amortized
over all future inference query services. The resulting profiled data is stored
as a two-dimensional lookup table that is indexed using (GPU partition
		size, batch size) which returns the (profiled) DNN execution time. Because
the lookup table separately keeps track of the profiled DNN execution time across
different GPU partition sizes, \elsa is able to accommodate the unique compute capabilities
of \paris's heterogeneous devices into its scheduling algorithm. Below we refer to
the estimated DNN execution time via our profiling-based lookup table as
\emph{T$_{estimated}$}.

\begin{figure}[t!] \centering
\includegraphics[width=0.485\textwidth]{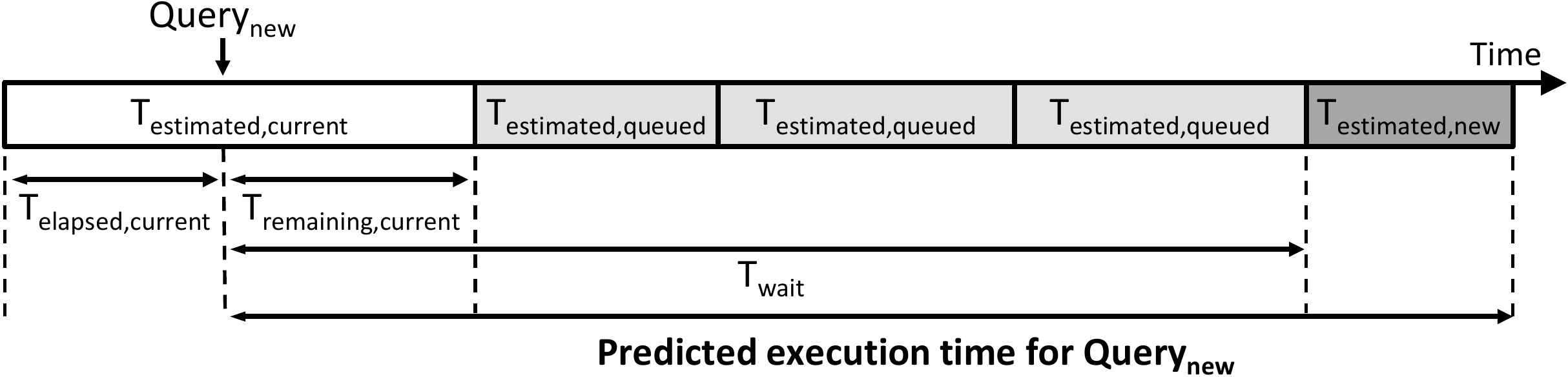}
\caption{
Estimating SLA slack for a newly arrived query when there are already multiple queries queued inside the server.
}
\label{fig:proposed_sla_slack_example}
\end{figure}

{\bf SLA slack time prediction.} Providing fast responsiveness is of highest
importance for end-users, so MLaaS providers have strict SLA targets to satisfy
to meet QoS requirements.  \elsa utilizes our DNN execution time estimator to
predict how much \emph{slack} a particular query has left until violating its SLA target (if any) over a
target GPU partition. \eqn{eqn:time_wait} and
\eqn{eqn:sla_slack} summarize our SLA slack prediction model:

\begin{equation}
\label{eqn:time_wait}
\footnotesize
T_{wait}=\Sigma{(T_{estimated,queued})} + T_{remaining,current}
\end{equation}

\begin{equation}
\label{eqn:sla_slack}
\footnotesize
SLA \ slack=SLA_{target}\ -\ \alpha(T_{wait} + \beta \cdot T_{estimated,new})
\end{equation}

 Whenever a \emph{new} service query is received at the server, \elsa first calculates
 how much time this new query must wait inside a target GPU partition
 until it gets a chance to be serviced (\emph{T$_{wait}$}, \eqn{eqn:time_wait}).  As depicted in
 \fig{fig:proposed_sla_slack_example}, all GPU partitions have its local scheduling queue
 that buffers all the queries yet to be executed by the GPU. Consequently,
 \emph{T$_{wait}$} can be estimated by calculating 1) the total amount of DNN
 model execution time expected to elapse when all the queries buffered inside
 the scheduling queue are fully executed
 ($\sum$(\emph{T$_{estimated,queued}$})), and 2) the remaining DNN model
 execution time of the query \emph{currently} being executed by the GPU
 (\emph{T$_{remaining,current}$}). Using our profile-based DNN execution time
 lookup table, \elsa can easily derive $\sum$(\emph{T$_{estimated,queued}$}).
		 As for \emph{T$_{remaining,current}$}, we employ a timestamp that starts
		 ticking whenever a new query starts execution on a GPU, which
		 we can utilize to measure how much execution time has elapsed since it started executing (\emph{T$_{elapsed,current}$} in \fig{fig:proposed_sla_slack_example}).
		 Because \emph{T$_{estimated,current}$}=(\emph{T$_{elapsed,current}$}+\emph{T$_{remaining,current}$}), 
		 \elsa uses the value of \emph{T$_{elapsed,current}$} to subtract it from the estimated end-to-end execution time
		 of the query currently executing on the GPU (\emph{T$_{estimated,current}$}) to derive
		 \emph{T$_{remaining,current}$}, allowing us to derive \emph{T$_{wait}$}.

 As the query's total wait time inside the server (\emph{T$_{wait}$})
	counts against SLA, our slack estimation model subtracts this amount from the
	model specific SLA target (\emph{SLA$_{target}$}). Additionally, the estimated DNN
	model execution time of the new query (\emph{T$_{estimated,new}$}) should also be accounted for when estimating the remaining
	SLA slack. As a result, \emph{T$_{estimated,new}$} is also subtracted from the SLA
	target to derive the final estimated SLA slack remaining for the new query (\eqn{eqn:sla_slack}).
	Note that $\alpha$ and $\beta$ are configurable parameters we employ to tune
	the SLA slack predictor's performance in accordance to the unique server
	environment \elsa is being deployed at.

\begin{algorithm}[t!]
\caption{ELSA}
\label{algo:elsa}
\begin{algorithmic}[1]
\scriptsize
\Procedure {Elastic\_Sched\_Algorithm()}{}
\Statex \LineComment{Step A: Schedule new query if the GPU partition can satisfy SLA}
\State $Sort \ GPU \ partitions \ in \ ascending \ order \ of \ partition \ size$
\For {$\textbf{each} \ GPU \ partition \ G$}
\If {$SLA > \alpha \times (T_{wait} + \beta \cdot T_{estimated,new}) $}
\State $Schedule \ query \ to \ GPU \ partition \ G$
\State \Return
\EndIf
\EndFor
\Statex
\Statex \LineComment{Step B: If Step A failed, schedule query to the GPU partition that can service the new query the fastest}
\State $t_{min} \leftarrow INT\_MAX$
\State $G_{min} \leftarrow -1$
\For {$\textbf{each} \ GPU \ partition \ G$}
\If {$t_{min} > T_{wait} + T_{estimated,new} $}
\State $t_{min} \leftarrow T_{wait} + T_{estimated,new}$
\State $G_{min} \leftarrow G$
\EndIf
\EndFor
\State $Schedule \ query \ to \ GPU \ partition \ G_{min}$
\State \Return
\EndProcedure
\end{algorithmic}
\end{algorithm}

{\bf Implementation.} With our SLA slack predictor in place, \elsa is able to
quantify which among the heterogeneously partitioned GPUs are able to service
the subject query without SLA violations (if it is at all possible).
\algo{algo:elsa} is a pseudo-code of \elsa, which goes through two primary
steps.  During the first step, we iterate through all available GPU partitions
and calculate the SLA slack had the subject query been scheduled to the subject
GPU partition (line $2$-$9$). Note that our scheduling algorithm iterates
through the smaller GPU partitions first (line $3$-$4$), prioritizing the
scheduling of new queries to smaller GPU partitions if there are multiple GPU
partitions that satisfy SLA (line $5$-$7$). This is because,
assuming the SLA slack is large enough, servicing a query using
a smaller GPU partition is \emph{always} beneficial from a GPU utilization
perspective, i.e., if the same query is executed on a larger GPU, it is
likely that the resulting GPU utilization will be lower than what it would
have been had it executed on a
smaller GPU. 

In the case where none of the GPU partitions are currently able to satisfy SLA
for the new query, we schedule this query to a GPU partition that will take the
\emph{least} amount of service time (line $10$-$21$). As the chances of this
new query to fulfill SLA is low, we empirically find that minimizing its
presence inside the inference server (i.e., by servicing it as quickly as possible) also
minimizes the deteriorating effects it has on \emph{other} queries that can
still satisfy SLA.

\begin{figure}[t!] 
\centering
\subfloat[FIFS]{
\includegraphics[width=0.42\textwidth]{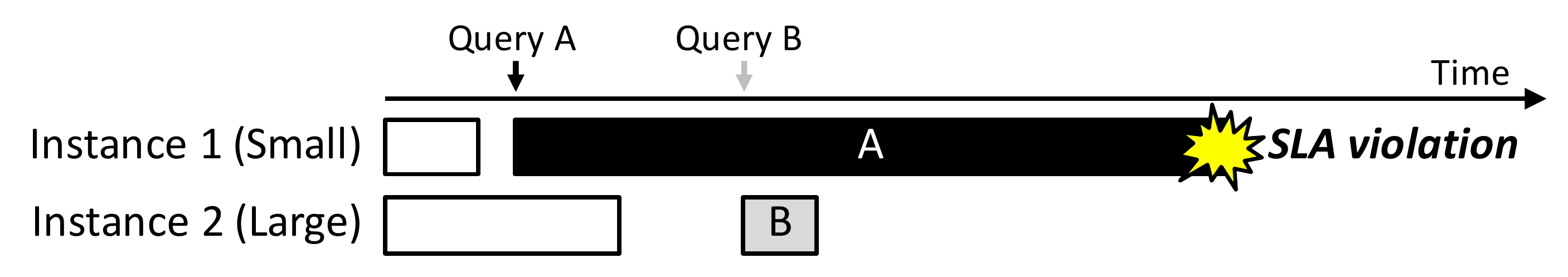}
\vspace{-1.5em}
	\label{fig:}
}
\vspace{-1em}
\subfloat[\elsa]{
	\includegraphics[width=0.42\textwidth]{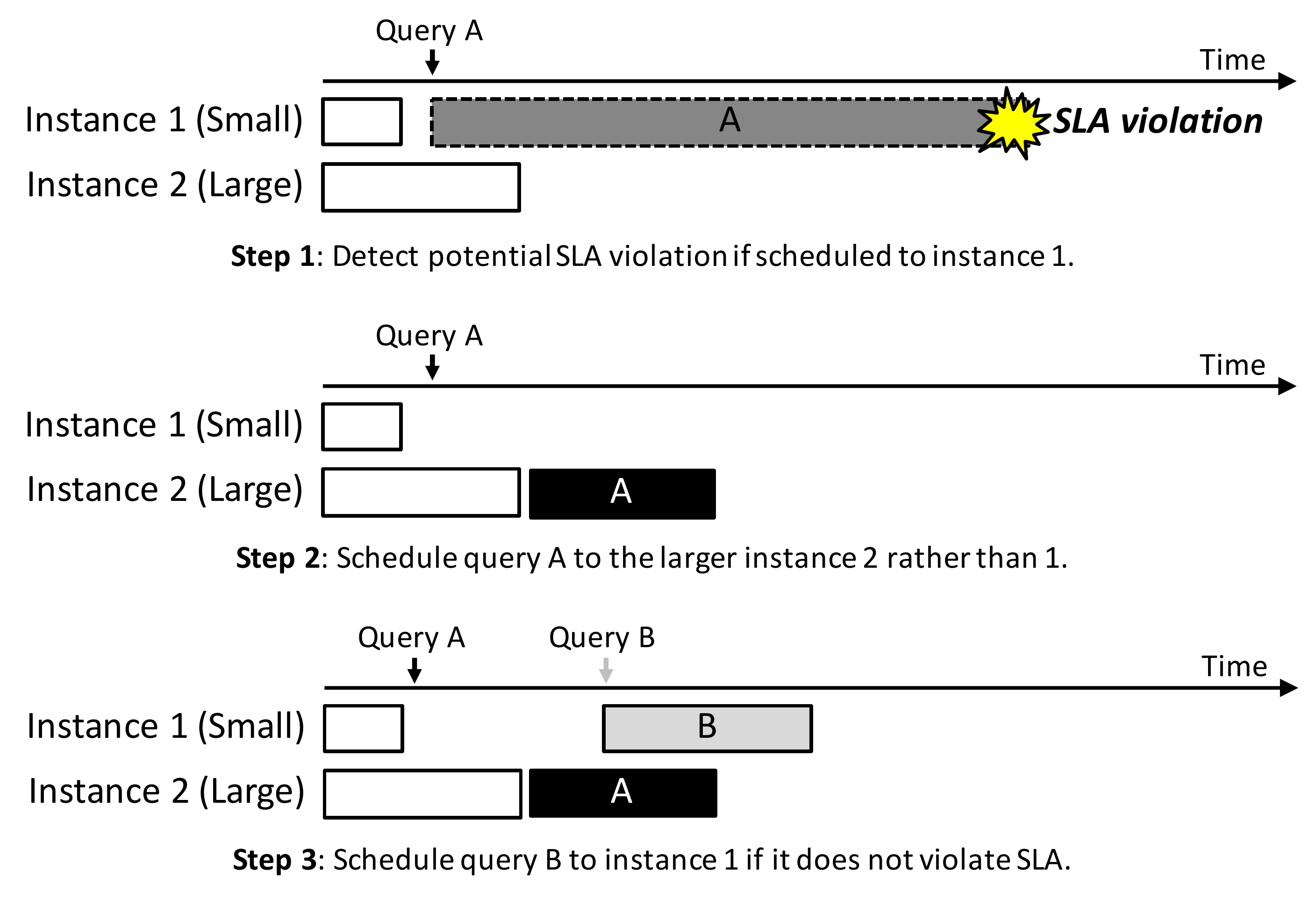}
	\vspace{-1.5em}
	\label{fig:}
}
\caption{ 
Timeline of how the two queries $A$ and $B$ are handled when using (a) FIFS and (b) \elsa.
}   
\label{fig:proposed_elsa}
\end{figure}

\fig{fig:proposed_elsa} provides an illustrative example on the advantages of our
heterogeneity-aware \elsa vs. FIFS. As depicted, FIFS fails to realize that
query $A$ can lead to significantly longer latency when executed on the small GPU
partition, thus violating SLA. In contrast, \elsa uses our SLA slack predictor to
acknowledge the potential of such hazardous situation and instead decides to 
schedule this query to the large GPU partition, avoiding SLA violations.

\section{Methodology}
\label{sect:methodology}

{\bf Benchmarks.} We study five DNN models 
used for computer vision
(ShuffleNet~\cite{shufflenet}, MobileNet~\cite{mobilenet}, ResNet~\cite{resnet}), natural language
processing (BERT~\cite{bert}), and automatic speech recognition (Conformer~\cite{conformer}). We chose these workloads as they exhibit
different levels of compute-intensity (i.e., low (ShuffleNet, MobileNet), medium (ResNet, Conformer),
		and high (BERT)), thus enabling us to explore the sensitivity of \paris and
\elsa under diverse	DNN model's unique compute/memory requirements.

{\bf Query size distribution, query arrival rate.} The size of a query
determines the input batch size for an inference. Prior
work~\cite{deeprecsys,li2016work,barford1998generating} observes that the
query size distribution follows a log-normal distribution. Therefore, we model
our batch size
distribution PDF to follow a log-normal distribution with a batch
size ranging from $1$ to $32$ in our default configuration. In terms of query arrival rates,
we employ MLPerf inference benchmark's
recommended Poisson distribution for modeling the rate at which a new query
arrives to the inference server.
In \sect{sect:sensitivity}, we
evaluate the sensitivity of \paris and \elsa under different batch size
distributions. 

{\bf Software.} We implemented the software runtime system of our multi-GPU
inference server by heavily modifying Facebook's open-sourced
DeepRecInfra~\cite{deeprecsys_github}, a software framework that enables the
modeling of at-scale datacenter environment's query size distribution, query
arrival rates, and etc (which is configured as discussed above). 
Our ML inference server is setup on top of Ubuntu $18.04$ and PyTorch $1.7.1$
backed with CUDA $11.1$ and cuDNN $8.0$.

{\bf Hardware.} We conduct our experiments on an Amazon EC2 p4d instance
(\emph{p4d.24xlarge}), which contains $8$ NVIDIA A100 GPUs, $96$ vCPUs, and
$1152$ GBs of host memory.  As each A100 contains $7$ GPCs
(\sect{sect:gpu_arch}), a max total of ($7\times8$)=$56$ GPCs can be utilized
by \paris to allocate the appropriate number of GPCs per each GPU partition and
instantiate them in our inference server.  Note that configuring
a homogeneously partitioned multi-GPU server faces several challenges under
some of the GPU partition granularities because of the odd-numbered $7$ GPCs
available per each A100 GPU.  For instance, when seeking to configure a
homogeneous server with \Medium{4}, a single A100 can only instantiate one
instance of \Medium{4} and must leave the remaining 3 GPCs idle.  Consequently,
				 the evaluation section (\sect{sect:eval}) primarily focuses on
				 \Small{1,2}/\Medium{3}/\Largest{7} as the partitioning granularity
				 when studying homogeneous servers configured using small/medium/large
				 sized GPUs, respectively. Below we detail how the number of instances
				 for each GPU partitions is configured for homogeneous and
				 heterogeneous servers.

\begin{table*}[t!]
\centering
\caption{The set of homogeneous vs. heterogeneous GPU partition configurations we explore in \sect{sect:eval}.}
\scriptsize
\begin{tabular}{|c|c|c|c|c|c|c|c|c|c|c|}
\hline
\multirow{2}{*}{\textbf{}} & \multicolumn{2}{c|}{\textbf{ShuffleNet}} & \multicolumn{2}{c|}{\textbf{MobileNet}} & \multicolumn{2}{c|}{\textbf{ResNet}}                       & \multicolumn{2}{c|}{\textbf{BERT}}                          & \multicolumn{2}{c|}{\textbf{Conformer}}                            \\ \cline{2-11}
& \begin{tabular}[c]{@{}c@{}}\#instance\end{tabular} & \#GPC & \begin{tabular}[c]{@{}c@{}}\#instance\end{tabular} & \#GPC & \begin{tabular}[c]{@{}c@{}}\#instance\end{tabular} & \#GPC & \begin{tabular}[c]{@{}c@{}}\#instance\end{tabular} & \#GPC & \begin{tabular}[c]{@{}c@{}}\#instance\end{tabular} & \#GPC \\ \hline

\hline
\Small{1}   & 24 & 24 & 24 & 24    & 48       & 48    & 42  & 42 & 48 & 48    \\ \hline
\Small{2}   & 12 & 24 & 12 & 24    & 24       & 48    & 21  & 42 & 24 & 48   \\ \hline
\Medium{3}  & 8 & 24 & 8  & 24    & 16       & 48    & 14  & 42  & 16 & 48  \\ \hline
\Largest{7} & 4 & 28 & 4  & 28    & 8        & 56    & 6   & 42 & 8 & 56   \\ \hline
\hline
Random			& varies & 24 & varies  & 24    & varies        & 48    & varies   & 42 & varies & 48   \\ \hline
\hline
\paris			& varies & 24 & varies  & 24    & varies        & 48    & varies   & 42 & varies & 48   \\ \hline
\hline
\# of A100                 & \multicolumn{2}{c|}{4}   & \multicolumn{2}{c|}{4}                                     & \multicolumn{2}{c|}{8}                                        & \multicolumn{2}{c|}{6} & \multicolumn{2}{c|}{8}                                       \\ \hline
\end{tabular}
\label{tab:partition}
\end{table*}

\begin{figure*}[t!] 
\centering
\includegraphics[width=0.91\textwidth]{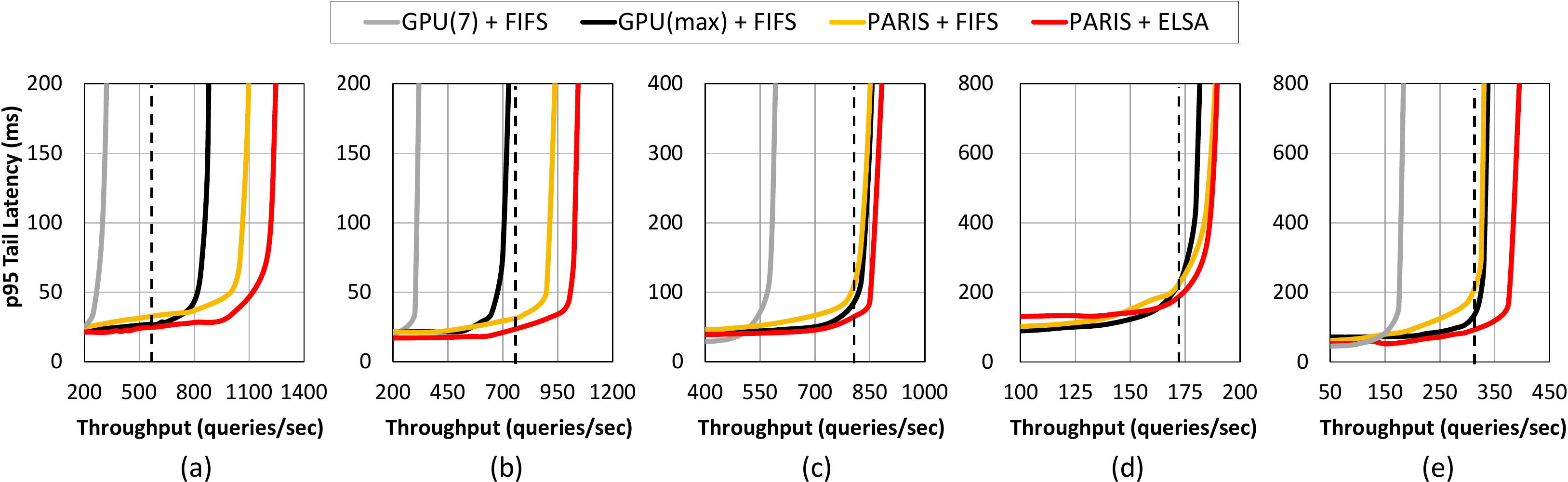}
\caption{ 
$95$-percentile tail latency (y-axis) and latency-bounded throughput (i.e., the number of queries processed per second that meets a target tail latency, x-axis) for (a) ShuffleNet, (b) MobileNet, (c) ResNet, (d) BERT, and (e) Conformer. For brevity, we only plot GPU(7) and GPU(max) as these two designs provide best average latency-bounded throughput among all baseline designs we study (detailed in \fig{fig:eval_throughput}). (d) BERT does not show GPU(7)+FIFS because GPU(max) equals GPU(7).
}   
\label{fig:eval_latency}
\end{figure*}

{\bf Configuration of homogeneous vs. heterogeneous GPU partitions.}

\tab{tab:partition} summarizes our
				 studied server configurations for the five DNN models.
				 There are several things worth clarifying in our evaluation settings and we
				 use the configurations of MobileNet/ResNet/BERT to highlight these points.
				 First, in most of our experiments, we were not able to fully utilize
				 the max $56$ GPCs because of the limited number of ways we can
				 practically partition the A100 GPUs while allowing all homogeneous
				 \Small{1,2}/\Medium{3}/\Largest{7} based servers to use the same
				 number of GPCs (e.g., $56$ and $28$ GPCs cannot be evenly divided with
				 \Medium{3}).  Second, note how the total number of GPCs utilized
				 for MobileNet is smaller than those used for ResNet/BERT. We observe
				 that MobileNet's (relatively) short DNN execution time makes the
				 ``total of $48$ GPCs, $48$ instances of \Small{1}'' design point to
				 become completely bottlenecked by the frontend of the inference server
				 (which supplies input queries to the GPUs) because the backend GPU
				 workers consume the incoming queries at a much higher throughput than
				 the queries supplied to the GPUs. Such unbalanced system design point
				 defeats the purpose of comparing different homogeneously partitioned
				 server design points vs. our proposal.  Therefore, MobileNet is
				 studied with max $24$ GPCs (a design point that all homogeneous servers
				 do not experience the aforementioned frontend bottleneck) for all homogeneous server
				 configurations (with the exception of \Largest{7}) as well as \paris. Because
				 the max $24$ GPCs in MobileNet cannot be evenly partitioned using \Largest{7}, we
				 employ the closest number $28$ GPCs ($4$ instances of \Largest{7}) as
				 the homogeneously partitioned large GPU server. Same principle holds
				 for ResNet's \Largest{7} setting, where we employ $8$ instances of
				 \Largest{7} (total $56$ GPCs) vs. the total $48$ GPCs used under
				 \Small{1,2,3}. Because all of our \paris design points are given only
				 $24$/$48$/$42$ GPCs for MobileNet/ResNet/BERT as the pool of GPCs for
				 partitioning (i.e., identical to the number of GPC given to \Small{1,2,3} and
						 smaller than the total number of GPCs assigned to
						 \Largest{7}), our evaluation provides a conservative
				 estimation of the benefits provided with \paris and \elsa. While these
				 caveats might give the impression that the usage of reconfigurable
				 GPUs are limited, recall that A100 is the first GPU to employ
				 reconfigurability, so we expect these issues to be resolved in future
				 iterations of GPUs.

{\bf SLA target.} As the precise SLA target numbers per each DNN model are vendor-specific,
	proprietary information not publicly disclosed, we take the
	following measure in setting our SLA target when measuring tail latency. For
	a given query size distribution, we first measure the DNN model's inference latency with 
	the distribution's max
	batch size ($32$ under our default setting) over \Largest{7}. The SLA target for a given model is setup
	as $N$ times (=$1.5\times$ in our default setting) larger than this measured inference latency. This is because the SLA should at least be large enough 
	for a given GPU partition handle. In \sect{sect:sensitivity}, we evaluate the sensitivity of our proposal to different SLA targets, i.e., different $N$ numbers.

\section{Evaluation} 
\label{sect:eval}

We compare the following six design points in our analysis:

\begin{enumerate}
\item $[$GPU(N)+FIFS$]$: homogeneous partitioning with GPU(N), (N: number of GPCs per GPU partition), schedule FIFS
\item $[$GPU(max)+FIFS$]$: GPU(max) reports the best performing homogeneous partitioning among all possible GPU(N), schedule FIFS
\item $[$Random+FIFS$]$: randomly partitioning the GPU in a heterogeneous manner, schedule with FIFS
\item $[$Random+ELSA$]$: randomly partitioning the GPU in a heterogeneous manner, schedule with ELSA
\item $[$PARIS+FIFS$]$: heterogeneous partitioning using PARIS, schedule with FIFS
\item $[$PARIS+ELSA$]$: heterogeneous partitioning using PARIS, schedule with ELSA
\end{enumerate}

Since there are many design points we explore in this section (e.g., GPU(N) alone contains four
		design points, N=1,2,3,7, \tab{tab:partition}), some of the
figures presented in this section do not show the results for all possible designs
for both brevity and clarity of explanation.  Specifically, we exclude
showing the results exhibiting low performance and use GPU(max) as an
optimistic homogeneous partitioning scheme (i.e., it performs as an
		upper bound design for homogeneous partitioning).
Note that we included ``Random'' partitioning as means to demonstrate the importance of accommodating
model properties and batch size distribution when heterogeneously partitioning the reconfigurable GPUs.

\subsection{Tail Latency}
\label{sect:eval_latency}
In \fig{fig:eval_latency}, we show latency-bounded throughput as a function of
a target tail latency. The vertical lines show the latency-bounded throughput
when the target tail latency is setup identically to our SLA. Using this as the
comparison point, the best performing homogeneous partition GPU(max)+FIFS
suffers from $1.1\times$, $17.4\times$, $1.2\times$, $1.1\times$, and $1.2\times$ worse tail latency than
PARIS+ELSA for ShuffleNet, MobileNet, ResNet, BERT, and Conformer, respectively. Here, MobileNet
performs best when configured with a GPU partition size of GPU(3)
	(i.e., GPU(max)=GPU(3), eight instances of GPU(3), $24$(=$3\times$$8$) GPCs overall,
	see \tab{tab:partition}). As discussed in \fig{fig:motivation_batch_effect},
	however, MobileNet suffers from significant GPU underutilization with a
	medium sized GPU(3) leaving significant room for improvement. Our
	PARIS observes such opportunity and utilizes the $24$ GPCs to construct
	a heterogeneous group of GPU partitions, specifically
	$6\times$GPU(1)+$4\times$GPU(2)+$2\times$GPU(3)+$1\times$GPU(4), which allows
	PARIS+ELSA to drastically improve tail latency and overall throughput
	(discussed in the next subsection). 

As for ResNet and BERT, these two models' GPU(max) is determined as GPU(3) and
GPU(7), respectively. Because the GPU underutilization under these GPU
partition sizes are not as significant under MobileNet, the tail latency
improvements with PARIS+ELSA is relatively modest compared to MobileNet.
Nonetheless, recall that GPU(max) is an optimistic, upper bound design point of a
homogeneously partitioned multi-GPU server. That is, determining the optimal GPU(max)
	design for homogeneous servers requires the system architect to
	painstakingly	 search through the wide design space in a manual,
	brute-force	manner. As discussed in \sect{sect:paris}, PARIS is a fully automated algorithm that
	systematically finds out the optimal partitioning points to pursue without
	any additional effort from the end-user.
\begin{figure}[t!] \centering
\includegraphics[width=0.485\textwidth]{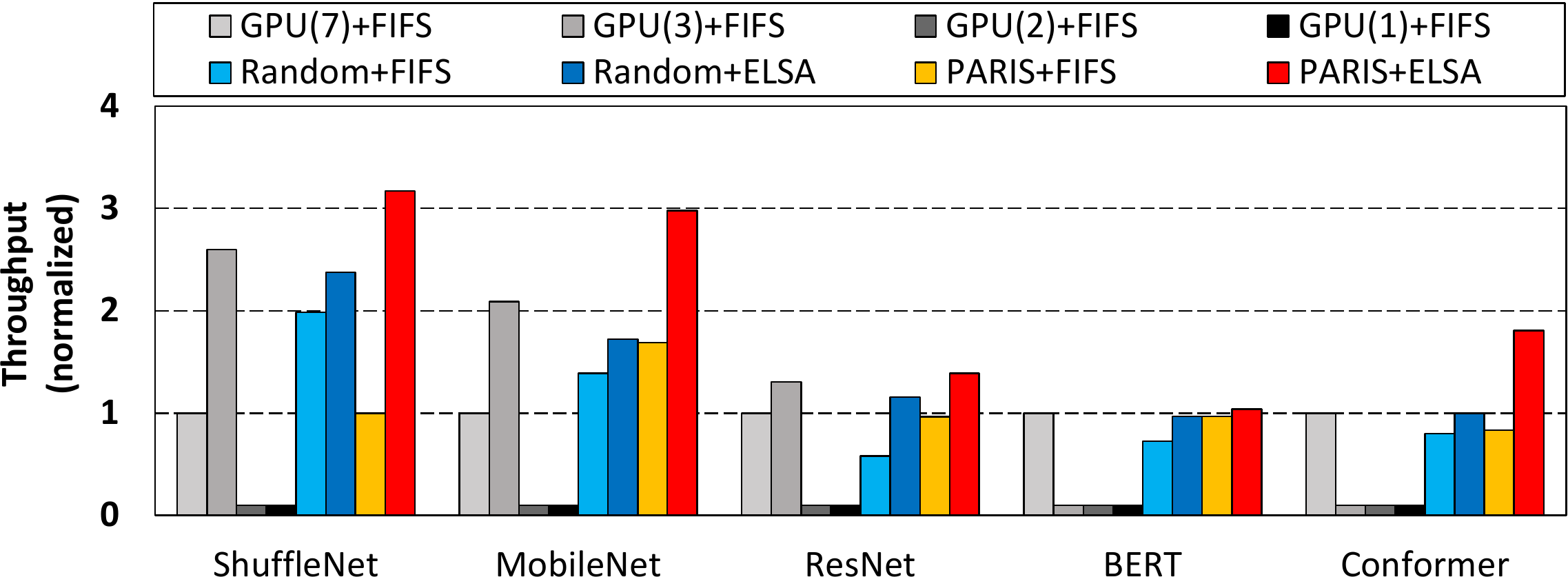}
\caption{
Latency-bounded throughput (normalized to GPU(7)+FIFS). The gray-black
	colored bars represent the homogeneously partitioned multi-GPUs.
}
\label{fig:eval_throughput}
\end{figure}

\subsection{Latency-bounded Throughput}
\label{sect:eval_throughput}

\fig{fig:eval_throughput} shows latency-bounded throughput, which is
normalized to GPU(7)+FIFS as it provides the most robust performance 
among all studied homogeneous server configurations.
	Several key observations can
	be made from this experiment. First, no single homogeneously partitioned
	GPU(N) design is able to universally achieve high throughput. For instance,
	GPU(3)+FIFS provides competitive results vs. our proposal for MobileNet ($70\%$ of PARIS+ELSA) and
	ResNet ($94\%$ of PARIS+ELSA). Unfortunately, GPU(3) suffers from substantial throughput degradation for
	BERT because it cannot provide enough computation power 
	to satisfactorily service this highly compute-intensive ML model. Consequently, GPU(3) suffers from significant SLA
	violations when BERT is executed with a large batch size, rendering GPU(7) the most
	robust design when considering all three models.
 PARIS, on the other hand, is able to identify the need for high computing power 
 within the inference server for BERT, 
	partitioning the $42$ GPCs (\tab{tab:partition}) into a heterogeneous group of
	$2\times$GPU(3)+$2\times$GPU(4)+$4\times$GPU(7). Such heterogeneity allows our proposed inference
	server to flexibly adapt to the unique DNN computation demands of BERT. 

		Another important point worth mentioning is the effectiveness of our \elsa
	algorithm, especially for MobileNet and ResNet.  
	Take MobileNet as an example, which \paris configures the $24$
	GPCs into $6\times$GPU(1)+$4\times$GPU(2)+$2\times$GPU(3)+$1\times$GPU(4).
	Because the number of large GPUs \paris has provisioned is relatively small,
	the scheduler should schedule large batch queries judiciously in order to
	minimize SLA violations. The heterogeneity-aware \elsa utilizes our SLA slack
	estimator to predict the likelihood of SLA violations and does a better
	job handling large batch queries than FIFS, providing high throughput improvements.  
	BERT is the least sensitive to the addition of \elsa in \paris, as
	PARIS+FIFS already provides superior performance, leaving little rooms of improvement. 

		It is interesting to note that a randomly partitioned heterogeneous server
	performs fairly competitively vs. homogeneous servers, provided it is coupled
	with our \elsa scheduler. These results highlight the merits of adding
	heterogeneous compute capabilities into ML inference servers.

	Overall, our fully automated PARIS+ELSA demonstrates the importance of
	incorporating heterogeneity into reconfigurable multi-GPU servers tailored for ML inference.

\subsection{Sensitivity}
\label{sect:sensitivity}

\begin{figure}[t!] 
\centering
\subfloat[]{
\includegraphics[width=0.485\textwidth]{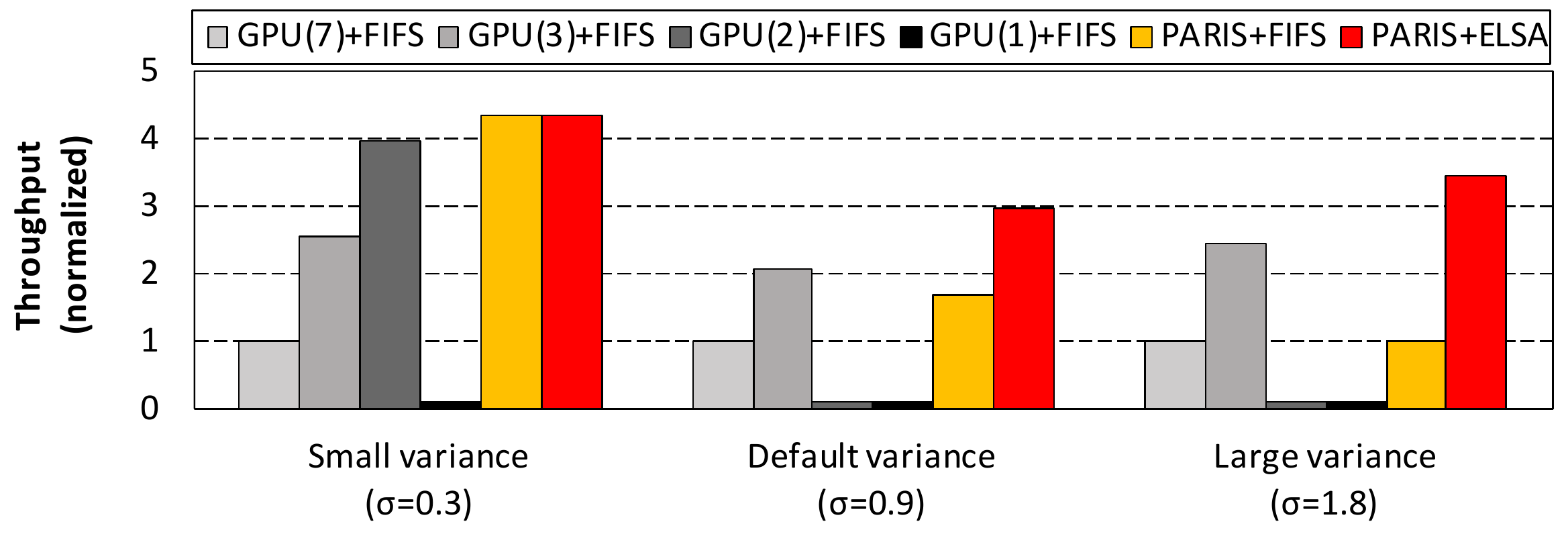}
\vspace{-1.5em}
	\label{fig:}
}
\vspace{-1em}
\subfloat[]{
\includegraphics[width=0.485\textwidth]{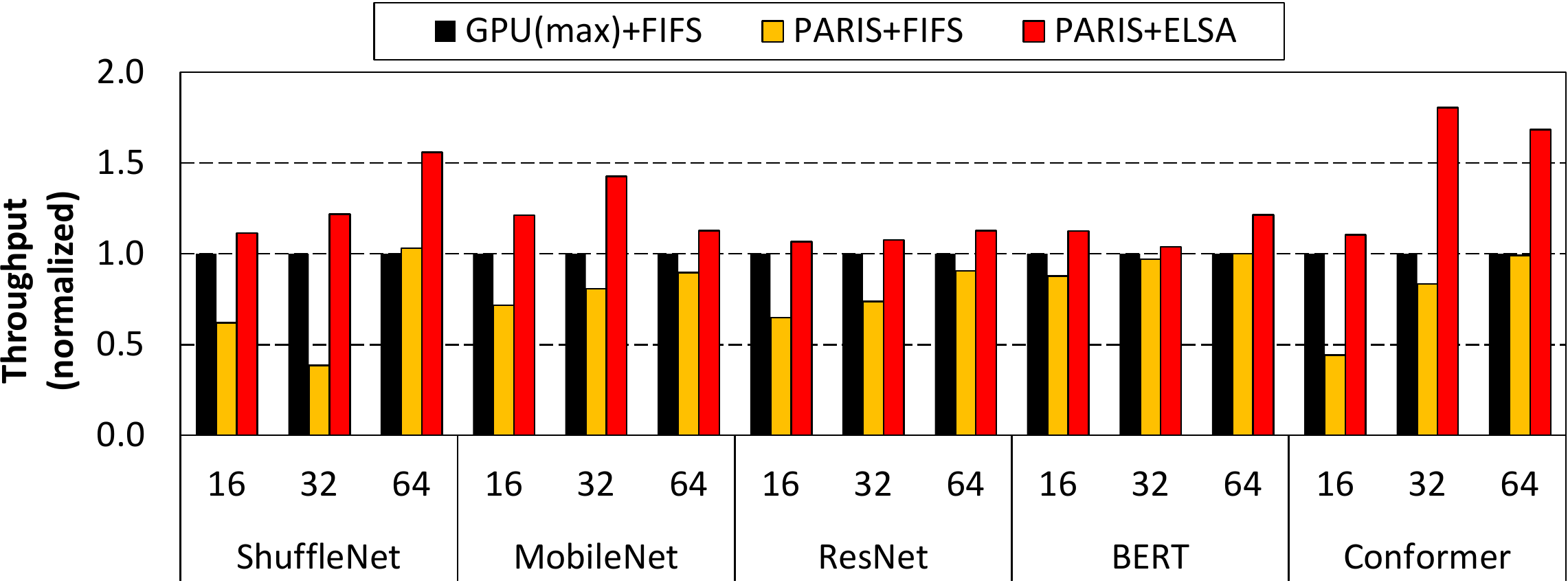}
	\vspace{-1.5em}
	\label{fig:}
}
\caption{ 
PARIS+ELSA sensitivity to (a) different log-normal distribution parameters and (b) different maximum batch size within the distribution.
}   
\label{fig:eval_sensitivity}
\end{figure}
{\bf Batch size distribution.} \fig{fig:eval_sensitivity}(a) summarizes the
sensitivity of our proposal to different log-normal distributions, i.e., when
changing the distribution variance values from small to large.  Under small
variance distributions, the rooms of improvement a heterogeneous
multi-GPU server can fulfill are relatively smaller. This is because under small(er)
variance log-normal distributions, the batch sizes tend to be centered around
a specific value which gives more likelihood of a specific homogeneous partitioning point
to more robustly handle inference queries. Consequently, 
the throughput
improvements provided with PARIS+ELSA compared to the best performing GPU(max)
	become smaller (larger) with smaller (larger) variance distributions. 

{\bf Max batch size.} \fig{fig:eval_sensitivity}(b) shows the throughput when the maximum batch size within our
batch size distribution is changed. As depicted, the efficacy of PARIS+ELSA remains robust across wide ranging
max batch sizes.

{\bf Different SLA targets.} We also confirmed PARIS+ELSA's robustness under different SLA targets. For instance,  
	when the SLA target is setup as $N$(=$2.0\times$) times of the max batch size inference latency ($N$=$1.5\times$ being our
			default, \sect{sect:methodology}), PARIS+ELSA provides an average $1.19\times$ reduction in tail latency
	which translates into an average $1.7\times$ and $1.1\times$ improvement in latency-bounded throughput
	vs. GPU(7) and GPU(max), respectively.

\section{Conclusion}
\label{sect:conclusion}

We explore an emerging reconfigurable GPU architecture 
to construct a heterogeneous ML inference server. We first proposed
\paris, a partitioning algorithm for reconfigurable GPUs that systematically
determines a heterogeneous set of multi-granular GPU partitions in a user-transparent
manner. The heterogeneously partitioned multi-GPU server is orchestrated by 
\elsa, which is capable of exploiting the unique heterogeneous
computing power of \paris inference server for maximum efficiency. 
PARIS and ELSA require no additional effort from the end-user and provides 
high server utilization improvements while guaranteeing SLA.

\bibliographystyle{ieeetr}
\bibliography{references}

\end{document}